\documentclass[prb,aps,superscriptaddress,eqsecnum,twocolumn,amssymb,showpacs]{revtex4-1}

\usepackage{color}
\usepackage{graphicx}
\usepackage{longtable}
\usepackage{epsfig}
\usepackage{dcolumn}
\usepackage{bm}
\usepackage{amssymb}
\usepackage{epsfig}
\usepackage{graphicx}
\usepackage{dcolumn}
\usepackage{bm}

\begin{document}

\title{ Exotic spin orders driven by orbital fluctuations in the Kugel-Khomskii model }

\author{Wojciech Brzezicki}
\affiliation{Marian Smoluchowski Institute of Physics, Jagellonian University,
Reymonta 4, 30-059 Krak\'ow, Poland }

\author{Jacek Dziarmaga}
\affiliation{Marian Smoluchowski Institute of Physics, Jagellonian University,
Reymonta 4, 30-059 Krak\'ow, Poland }

\author{Andrzej M. Ole\'{s} }
\affiliation{Marian Smoluchowski Institute of Physics, Jagellonian University,
Reymonta 4, 30-059 Krak\'ow, Poland }
\affiliation{Max-Planck-Institut f\"ur Festk\"orperforschung, Heisenbergstrasse 1,
D-70569 Stuttgart, Germany }

\date{8 November 2012}
\begin{abstract}
We study zero temperature phase diagram of the three-dimensional
Kugel-Khomskii model on a cubic lattice using the cluster mean field
theory and different perturbative expansions in the orbital sector.
The phase diagram is rich, goes beyond the single-site mean field
theory due to spin-orbital entanglement. In addition to the
antiferromagnetic (AF) and ferromagnetic (FM) phases, one finds also a
plaquette valence-bond phase with singlets ordered either on horizontal
or vertical bonds. More importantly, for increasing Hund's exchange we
identify three phases with exotic magnetic order stabilized by orbital
fluctuations in between the AF and FM order:
(i) an AF phase with two mutually orthogonal
antiferromagnets on two sublattices in each $ab$ plane and
AF order along the $c$ axis (ortho-$G$-type phase),
(ii) a canted-$A$-type AF phase with a non-trivial canting angle between
nearest neighbor FM layers along the $c$ axis, and
(iii) a striped-AF phase with anisotropic AF order in the $ab$ planes.
We elucidate the mechanism responsible for each of the above phases
by deriving effective spin models which involve second and third
neighbor Heisenberg interactions as well as four-site spin interactions
going beyond Heisenberg physics, and explain how the entangled nearest
neighbor spin-orbital superexchange generates spin interactions between
more distant spins.
\\ \hskip .4cm
{\it Published in Physical Review B} {\bf 87}, 064407 (2013).
\end{abstract}


\maketitle

\section{Introduction}\label{intro}

Recent interest and progress in the theory of spin-orbital superexchange
models was triggered by the observation that orbital degeneracy
drastically increases quantum fluctuations (QF) which may suppress
long-range order in the regime of strong competition between different
types of ordered states near the quantum critical point.\cite{Fei97} The
simplest and archetypical three-dimensional (3D) model for the
spin-orbital physics is the Kugel--Khomskii (KK) model introduced for
KCuF$_{3}$ by Kugel and Khomskii long ago.\cite{Kug72,*Kug82} KCuF$_{3}$
is a strongly correlated system with a single hole within degenerate
$e_{g}$ orbitals at each Cu$^{2+}$ ion. Kugel and Khomskii showed that
interactions between strongly correlated electrons could then stabilize
orbital order by a purely electronic mechanism, the superexchange.
A similar situation occurs in a number of compounds with active orbital
degrees of freedom, where strong on-site Coulomb interactions localize
electrons (or holes) and give rise to spin-orbital superexchange.
\cite{Nag00,Hfm,Kha05,Ole05,Ole12}

The orbital superexchange may stabilize the orbital order by itself, but
in $e_{g}$ systems it is usually helped by the Jahn-Teller distortions
of the lattice which generate effective intersite orbital interactions.
\cite{Zaa93,Fei99,vdB99} For instance, in LaMnO$_{3}$ the terms
that originate from the superexchange and the Jahn-Teller
distortions are of equal importance and both are necessary to explain
the observed high temperature $T_{\rm OO}\simeq 780K$ of the structural
transition.\cite{Fei99} Also in KCuF$_{3}$ the lattice distortions play
an important role \cite{Pao02,*Cac02,*Bin04,Dei08,Leo10} and are
responsible for its strongly anisotropic magnetic and optical properties.
The latest theoretical and experimental results for this compound show
that other types of interactions, like Goodenough processes in the
superexchange,\cite{Ole05} direct orbital exchange driven by a
combination of electron--electron interactions and ligand distortions,
\cite{Lee12} or dynamical Dzyaloshinsky-Moriya interaction,\cite{Ere08}
are necessary to explain structural phase transition in KCuF$_{3}$ at
$T_{\rm OO}\simeq 800K$ and peculiarities in spin dynamics.\cite{Yam94}
The compound itself is believed to be the best realization of the
one-dimensional (1D) antiferromagnetic (AF) Heisenberg model above the
N\'eel temperature $T_{N}=39K$,\cite{Lak05} and spinon excitations were
indeed observed in neutron scattering.\cite{Ten93}

While the coexisting $A$-type AF ($A$-AF) order and the orbital order
are well established in KCuF$_{3}$ below the N\'eel temperature
$T_{N}\simeq 39$ K,\cite{Lee12} and this phase is reproduced by the
spin-orbital $d^{9}$ superexchange model in the mean field (MF)
approximation,\cite{Ole00} the phase diagram of this model is still
unknown beyond the MF approach because of strongly coupled spin and
orbital degrees of freedom\cite{Fei97,Fei98} which poses an outstanding
question in the theory: Which types of coexisting spin and orbital order
(or disorder) are possible when its microscopic parameters (splitting of
the $e_g$ orbitals $E_z$ and Hund's exchange $J_H$) are varied? So far,
it was suggested that the long-range AF order is destroyed by spin-orbital
QF,\cite{Fei97,Fei98} but another possibility that the
ordered state might be stabilized by the order-out-of-disorder mechanism
was also pointed out in the regime of ferro-orbital (FO) order of
$3z^2-r^2$ orbitals.\cite{Kha97} An alternative to magnetic order are
spin-disordered phases with pronounced valence-bond spin-orbital
correlations, as suggested by simple variational wave functions.\cite{Fei97}

The purpose of this paper is to investigate the phase diagram of the 3D
KK model, in the presence of spin-orbital QF. This
subject is of interest in the broad context of frustration in magnetic
systems,\cite{Nor09,Bal10} which appears to be intrinsic for the orbital
superexchange.\cite{Fei97,Cin10}
To establish reliable results concerning short-range order
in the crossover regime between phases with long-range AF or
ferromagnetic (FM) order, we developed a \textit{cluster MF approach\/}
which goes beyond the single-site MF in the spin-orbital system
\cite{DeS03} and is based on an exact diagonalization of a cluster
coupled to its neighbors by the MF terms. The cluster is chosen to be
sufficient for investigating both AF phases with four sublattices and
valence-bond states, with spin singlets either along the $c$ axis or
within the $ab$ planes. This theoretical approach is motivated by
possible spin-orbital entanglement\cite{Ole06,Ole12} which is
particularly pronounced in the 1D SU(4) [or SU(2)$\otimes$SU(2)]
spin-orbital model\cite{Fri99,*Ole07,*You12} and occurs also on the
frustrated triangular lattice,\cite{Nor08,*Nor11,*Cha11,*Tro12} and in
the models for perovskites when spin correlations are AF on the bonds
\cite{Ole12} --- then the Goodenough-Kanamori rules\cite{Goode,*Kan59}
are violated in some cases.
In the perovskite vanadates such entangled states play an important role
at finite temperature: in their optical properties,\cite{Kha04} in the
phase diagram,\cite{Hor08} and in the dimerization of FM interactions
along the $c$ axis in the $C$-AF phase of YVO$_{3}$,\cite{Ulr03,*Hor03}
understood within the 1D spin-orbital model.\cite{Sir08,*Sir03,*Her11}
Previous studies employing the cluster MF approach have shown that
phases with entangled spin-orbital degrees of freedom occur in the KK
bilayer,\cite{wb11,*cam11} while a noncollinear spin order emerges from
entangled spin-orbital fluctuations in the two-dimensional (2D) KK
monolayer.\cite{wb12} Below we investigate whether spin-orbital
entangled states could also play a role in the present 3D KK model.

The paper is organized as follows. In Sec. \ref{sec:deriv} we briefly
introduce the spin-orbital superexchange in the KK model.
As the first approximation, in Sec. \ref{sub:1stmf_3d} we present the
phase diagram of the model obtained in the single-site MF approximation
--- this approach ignores any possible spin-orbital entanglement.
To include the effects of spin-orbital QF we introduce
the cluster MF method for the 3D system in Sec. \ref{sec:opa}, with two
different topologies of the cluster, and present the phase diagram
modified by spin-orbital entanglement in Sec. \ref{sec:opa}. It contains
three phases with exotic magnetic order: the ortho-$G$-AF phase similar
to that found recently for the monolayer,\cite{wb12} the canted-$A$-AF
phase, and the striped-AF phase.
Next we present the behavior of the order parameters, spin angles,
total magnetization, correlations and spin-orbital covariances for
two paths in the phase diagram which include certain exotic phases:
(i) from the $A$-AF through canted-$A$-AF to FM phase in Sec.
\ref{sub:From--AF-to}, and
(ii) from the striped-AF to $G$-AF phase in Sec.
\ref{sub:From-striped-AF-to}.
As we show in Sec. \ref{sec:orbif}, rather weak orbital fluctuations
are found in several phases and the orbital moments are reduced by them.
Then we explain and highlight the origin
of the exotic magnetic phases by deriving effective spin Hamiltonians
within perturbative expansion in the orbital sector. Spin models for the
ortho-G-AF phase, canted-$A$-AF phase, and striped-AF phase are derived
in Secs. \ref{sub:ortho}, \ref{sub:cant3d}, and \ref{sub:stripeHs},
respectively. In Sec. \ref{sub:3dcgaf}, using a similar perturbative
expansion, we explain the absence of the $C$-AF phase in the phase
diagram of the KK model obtained within the cluster MF approximation.
Summary and conclusions are given in Sec. \ref{sub:summa_3d}, while
certain additional details of the performed perturbative analysis are
presented in Appendices A-C.

\section{The Kugel--Khomskii model}
\label{sec:model}

\subsection{Frustrated spin-orbital superexchange}
\label{sec:deriv}

For realistic parameters the late transition metal oxides or fluorides
are strongly correlated and electrons localize in the $3d$ orbitals,
\cite{Ole87,*Grz91,Gra92} leading in cuprates to Cu$^{2+}$ ions with spin
$S=1/2$ in $d^9$ configuration and $e_g$ orbitals occupied by one hole:
\begin{eqnarray}
|x\rangle\equiv(x^{2}-y^{2})/\sqrt{2},\hskip.5cm
|z\rangle\equiv(3z^{2}-r^{2})/\sqrt{6}\,.\label{eq:eg}
\end{eqnarray}
The examples of such systems are:
KCuF$_{3}$ with 3D cubic lattice,
K$_{3}$Cu$_{2}$F$_{7}$ representing bilayer compounds, and
K$_{2}$CuF$_{4}$ and La$_{2}$CuO$_{4}$ with 2D square lattice.
We also use below the short-hand notation for the orbital basis,
$\{x,z\}$. Here $t_{2g}$ orbitals are split in the octahedral field and
do not couple to $e_{g}$'s by hopping through fluorine, so they can be
neglected. In what follows we investigate the 3D spin-orbital
superexchange model obtained by considering charge excitations between
transition metal ions,\cite{Ole00}
$d_i^9d_j^9\rightleftharpoons d_i^{8}d_j^{10}$ in the regime of large
$U$, and neglect the coupling to lattice distortions arising due to
the Jahn-Teller lattice distortions.

One finds the Heisenberg Hamiltonian for $S=\frac12$ spins coupled
to the orbital problem,
\begin{equation}
{\cal H}=
-\frac{1}{2}J\sum_{\gamma=a,b,c}\sum_{\langle ij\rangle||\gamma}
H_{ij}^{\gamma}-E_{z}\sum_{i}\tau_{i}^{c},
\label{eq:KKham}
\end{equation}
where bond-interaction terms $H_{ij}^{\gamma}$ are defined as
follows,
\begin{eqnarray}
H_{ij}^{\gamma} & = & \left(r_{1}\,\Pi_{t}^{(ij)}+r_{2}\,\Pi_{s}^{(ij)}\right)
\left(\frac{1}{4}-\tau_{i}^{\gamma}\tau_{j}^{\gamma}\right)\nonumber \\
& + & \left(r_{2}+r_{4}\right)\Pi_{s}^{(ij)}
\left(\frac{1}{2}-\tau_{i}^{\gamma}\right)
\left(\frac{1}{2}-\tau_{j}^{\gamma}\right).
\label{eq:Hij}
\end{eqnarray}
Here $\gamma=a,b,c$ labels the direction of a bond $\langle ij\rangle$
in the 3D system. The energy scale is given by the superexchange
constant,
\begin{equation}
J=\frac{4t^{2}}{U},
\label{eq:J}
\end{equation}
where the hopping $t$ is the effective intersite $(dd\sigma)$ hopping
element for an $e_g$ hole between $z$ orbitals along the $c$ axis,
\cite{Zaa93} and the other $e_g$ hopping elements obey the cubic
symmetry.\cite{Fei05} The orbital operators
at site $i$ are $\vec{\tau}_i=\{\tau_i^{a},\tau_i^{b},\tau_i^{c}\}$.
These operators represent $e_{g}$ orbital degrees of freedom, have
cubic symmetry in the 3D lattice and are expressed in terms of Pauli
matrices
$\{\sigma_{i}^{x},\sigma_{i}^{y},\sigma_{i}^{z}\}$ in the following way:
\begin{eqnarray}
\label{tau}
\tau_{i}^{a(b)}\equiv\frac{1}{4}(-\sigma_{i}^{z}\pm\sqrt{3}\sigma_{i}^{x}),
\hskip.5cm
\tau_{i}^{c}\equiv\frac{1}{2}\sigma_{i}^{z}.
\end{eqnarray}
The matrices $\{\sigma_{i}^{\gamma}\}$ act in the orbital space
(and have nothing to do with the physical spin ${\bf S}_{i}$ present
in this problem). Note that $\{\tau_{i}^{\gamma}\}$ operators are not
independent from one another because they satisfy the local constraint,
$\sum_{\gamma}\tau_{i}^{\gamma}\equiv0$.
The operators $\Pi_{ij}^{s}$ and $\Pi_{ij}^{t}$ stand for projections
of spin states on the bond $\langle ij\rangle$ on a singlet
($\Pi_{ij}^{s}$) and triplet ($\Pi_{ij}^{t}$) configuration,
respectively,
\begin{equation}
\Pi_{s}^{(ij)}=
\left(\frac{1}{4}-{\bf S}_{i}\cdot{\bf S}_{j}\right), \hskip .5cm
\Pi_{t}^{(ij)}=
\left(\frac{3}{4}+{\bf S}_{i}\cdot{\bf S}_{j}\right),
\label{eq:proje}
\end{equation}
for spins $S=\frac12$ at both sites $i$ and $j$ of the bond
$\langle ij\rangle$. Spin interactions obey the SU(2) symmetry. More
details on the derivation of Eq. (\ref{eq:Hij}) may be found in Ref.
\onlinecite{Ole00}.

The model Eq. (\ref{eq:KKham}) depends thus on two parameters:
\cite{Ole00}
(i) Hund's exchange coupling $\eta$, and
(ii) the crystal-field splitting of $e_g$ orbitals $E_{z}/J$.
The first of them is given by the ratio of Hund's exchange $J_{H}$
and intraorbital Coulomb element $U$ defined in a standard way as
in the degenerate Hubbard model,\cite{Ole83}
\begin{equation}
\eta\equiv\frac{J_{H}}{U},
\label{eq:eta}
\end{equation}
and determines the values of the coefficients in Eq. (\ref{eq:Hij}):
\begin{equation}
r_{1}=\frac{1}{1-3\eta},\hskip.5cm
r_{2}=\frac{1}{1-\eta},\hskip.5cm
r_{4}=\frac{1}{1+\eta}.
\end{equation}
The typical energies for the Coulomb $U$ and Hund's exchange $J_{H}$
can be deduced from the atomic spectra or from density
functional theory with constrained electron densities. Earlier studies
performed within the local density approximation (LDA) gave rather
large values of the interaction parameters for Cu$^{2+}$ ions:
\cite{Gra92} $U=8.96$ eV and $J_{H}=1.19$ eV. More recent studies used
the Coulomb interactions treated within the LDA+$U$ scheme and gave
somewhat reduced values:\cite{Lie95} $U=7.5$ eV and $J_{H}=0.9$ eV.
However, both parameter sets give a rather similar value of Hund's
exchange parameter $\eta$, being within the expected range
$0.10<\eta<0.15$ for strongly correlated late transition metal oxides.
\cite{Ole05} Note that the physically acceptable range is much broader,
i.e., $0<\eta<1/3$. The upper limit follows from the condition
$(U-3J_{H})>0$ for the high-spin excitation energy.

The last term of the $H_{e_{g}}$ Hamiltonian (\ref{eq:KKham}) lifts the
degeneracy of the two $e_{g}$ orbitals,
\begin{eqnarray}
H_{z}=-\frac{1}{2}E_{z}\sum_i(n_{ix}-n_{iz})\equiv
-E_{z}\sum_i\tau_{i}^{c}\,,
\label{eq:Hz}
\end{eqnarray}
where $\{n_{ix\sigma},n_{iz\sigma}\}$ are hole number operators in
$\{x,z\}$ orbitals (\ref{eq:eg}) at site $i$. It favors hole occupancy
of $x$ ($z$) orbitals when $E_{z}>0$ ($E_{z}<0$) and can be associated
with a uniaxial pressure along the $c$ axis or a crystal-field splitting
induced by a static Jahn-Teller effect.

\begin{figure}[t!]
\begin{minipage}[t]{0.5\columnwidth}%
\begin{center}\includegraphics[clip,width=3.93cm]{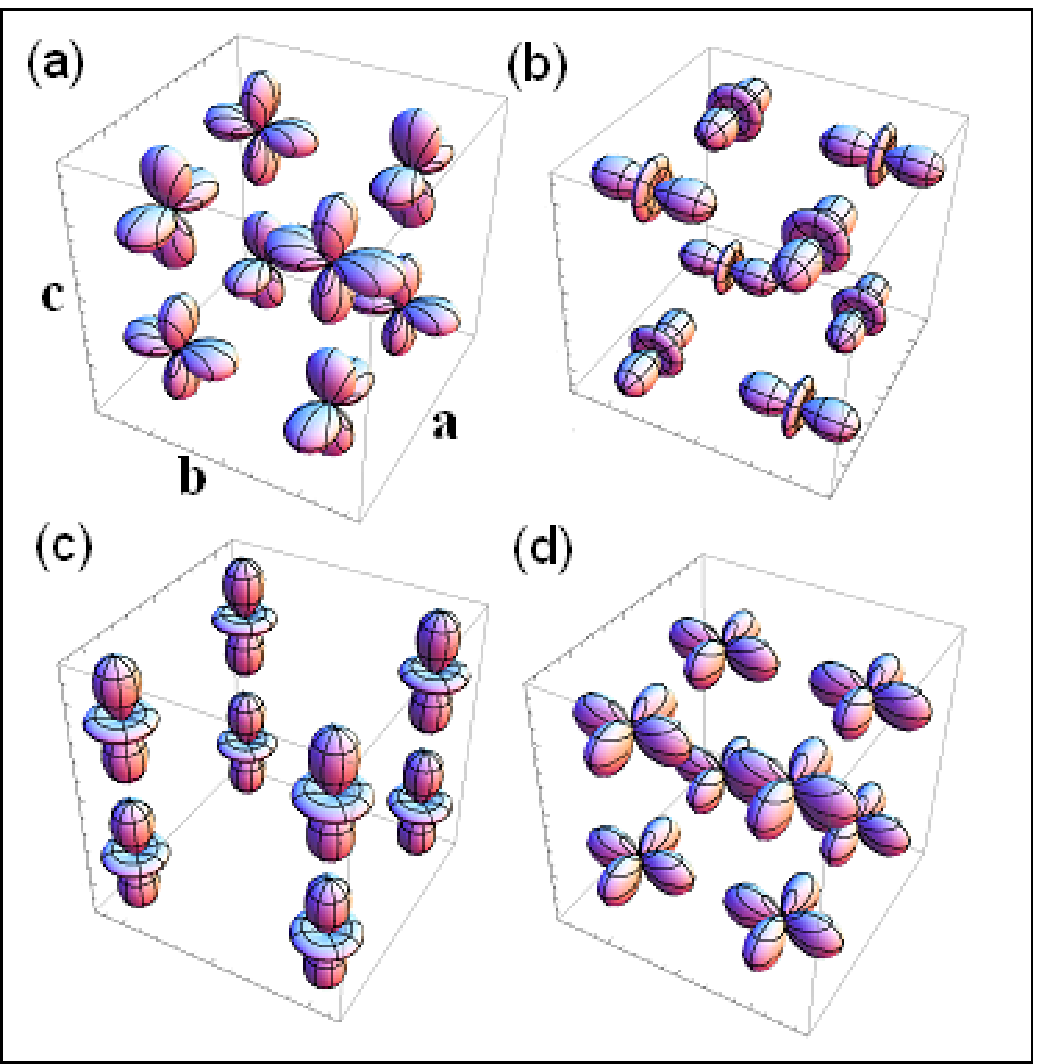}\end{center}%
\end{minipage}%
\begin{minipage}[t]{0.5\columnwidth}%
\begin{center}\includegraphics[clip,width=4.2cm]{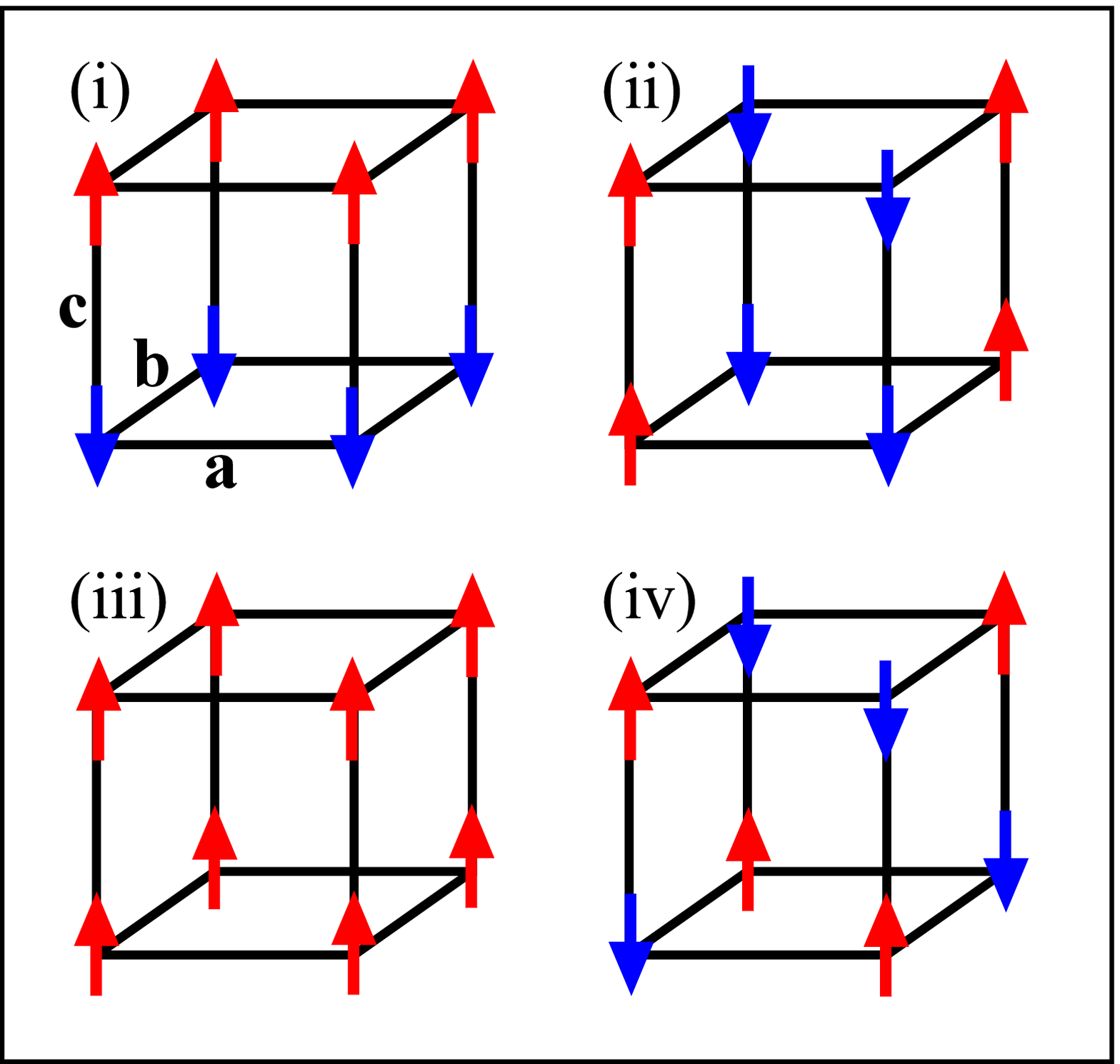}\end{center}%
\end{minipage}
\caption{(Color online)
Left panel: schematic view of four representative orbital configurations
on a representative cube of the 3D lattice:
(a) AO order with $\langle\tau_{i}^{a(b)}\rangle=1/2$ changing from site
to site and $\langle\tau_{i}^{c}\rangle=-1/4$, obtained for $E_{z}<0$,
(b) AO order with $\langle\tau_{i}^{a(b)}\rangle=-1/2$ changing from
site to site and $\langle\tau_{i}^{c}\rangle=1/4$, obtained for $E_{z}>0$,
(c) FO order with occupied $z$ orbitals and
$\langle\tau_{i}^{c}\rangle=-1/2$ (cigar-shaped orbitals), and
(d) FO order with occupied $x$ orbitals and
$\langle\tau_{i}^{c}\rangle=1/2$ (clover-shaped orbitals).
Right panel: schematic view of four representative spin configurations
(arrows stand for up or down spins) shown on a cubic cluster:
(i) $A$-AF configuration,
(ii) $C$-AF configuration,
(iii) FM configuration, and
(iv) $G$-AF configuration.
}
\label{fig:4cubes}
\end{figure}

In Figs. \ref{fig:4cubes}(a)-\ref{fig:4cubes}(d) we present typical
orbital configurations with FO order and alternating orbital (AO) order
considered in the $e_{g}$ orbital models.\cite{vdB99,Fei05}
In the next sections we analyze their possible
coexistence with spin order in the KK model Eq. (\ref{eq:KKham}).
As we can see, the maximal (minimal) value of the orbital operators
$\tau_{i}^{\gamma}$ is related with orbital taking shape of a clover
(cigar) with symmetry axis pointing along the direction $\gamma$.

\subsection{Phase diagram in single-site mean field}
\label{sub:1stmf_3d}

After averaging over spins, the Hamiltonian of Eq. (\ref{eq:KKham}),
originally expressed in terms of bond operators, can be rewritten as
an effective orbital Hamiltonian,
\begin{eqnarray}
{\cal H}_{{\rm MF}}\! & = & \frac{1}{2}J\sum_{i,\gamma}
\left\{ \tau_{i}^{\gamma}\tau_{i+\gamma}^{\gamma}(\chi^{\gamma}-\xi^{\gamma})
+\tau_{i}^{\gamma}\xi^{\gamma}
-\frac{1}{4}(\chi^{\gamma}+\xi^{\gamma})\right\} \nonumber \\
 & - & E_{z}\sum_{i}\tau_{i}^{c},
\end{eqnarray}
where $i$ runs over sites of the cubic lattice, and $i+\gamma$ is the
nearest neighbor (NN) of site $i$ along the axis $\gamma=a,b,c$.
The coefficients,
\begin{eqnarray}
\chi^{\gamma} = r_{1}\Pi_{t}^{\gamma}+r_{2}\Pi_{s}^{\gamma},\hskip .5cm
\xi^{\gamma}=(r_{2}+r_{4})\Pi_{s}^{\gamma},
\end{eqnarray}
are parameters obtained by averaging of the spin projectors in Eq.
(\ref{eq:proje}) under assumption that the spin order depends only on
the direction $\gamma$ and all the bonds $\langle i,i+\gamma\rangle$
along the axis $\gamma$ are equivalent:
\begin{equation}
\Pi_{s}^{\gamma}=
\frac{1}{4}-\langle{\bf S}_{i}\cdot{\bf S}_{i+\gamma}\rangle,
\hskip.5cm
\Pi_{t}^{\gamma}=
\frac{3}{4}+\langle{\bf S}_{i}\cdot{\bf S}_{i+\gamma}\rangle.
\label{eq:project}
\end{equation}
In the single-site MF the spin QF are absent at zero
temperature and the projectors can be replaced by their average values.
This is sufficient to investigate the phases with either AF or FM
long-range order.

\begin{table}[t!]
\caption{
Mean values of triplet $\Pi_{t}^{\gamma}$ and singlet $\Pi_{s}^{\gamma}$
projection operators Eqs. (\ref{eq:project}) for a bond
$\langle ij\rangle$ along the axis $\gamma=a,b,c$ in different phases
with long-range magnetic order which occur in the MF phase diagram,
see Figs. \ref{fig:4cubes}(i)-\ref{fig:4cubes}(iv).}
\begin{ruledtabular}
\begin{tabular}{ccccc}
 Phase & $\Pi_{t}^{a(b)}$ & $\Pi_{t}^{c}$ & $\Pi_{s}^{a(b)}$ & $\Pi_{s}^{c}$ \cr
\colrule
 $G$-AF & 1/2 & 1/2 & 1/2 & 1/2 \cr
 $C$-AF & 1/2 &  1  & 1/2 &  0  \cr
 $A$-AF &  1  & 1/2 &  0  & 1/2 \cr
    FM  &  1  &  1  &  0  &  0  \cr
\end{tabular}
\end{ruledtabular}
\label{tab:sord}
\end{table}

The values of the projection operators (\ref{eq:project}) depend on the
assumed spin order. Here we consider four different spin configurations
shown in Fig. \ref{fig:4cubes}:
(i) $A$-AF phase --- with FM order in the $ab$ planes and AF
correlations along the $c$ axis [Fig. \ref{fig:4cubes}(i)],
(ii) $C$-AF  phase --- with AF order in the $ab$ planes and FM
correlations along the $c$ axis [Fig. \ref{fig:4cubes}(ii)],
(iii) FM phase [Fig. \ref{fig:4cubes}(iii)], and
(iv) $G$-AF phase N\'eel state [Fig. \ref{fig:4cubes}(iv)].
In the single-site MF approximation we use the classical average values
of the spin projection operators in the above phases listed in Table
\ref{tab:sord}. Apart from fully AF and FM phase we include also $A$-AF
($C$-AF) configurations with spin correlations being AF along the $c$
axis (in the $ab$ planes), and FM otherwise. Solutions of the
self-consistency equations and ground state energies in different phases
can be obtained analytically, as shown in Ref. \onlinecite{wb11}.

\begin{figure}[b!]
\includegraphics[clip,width=8.4cm]{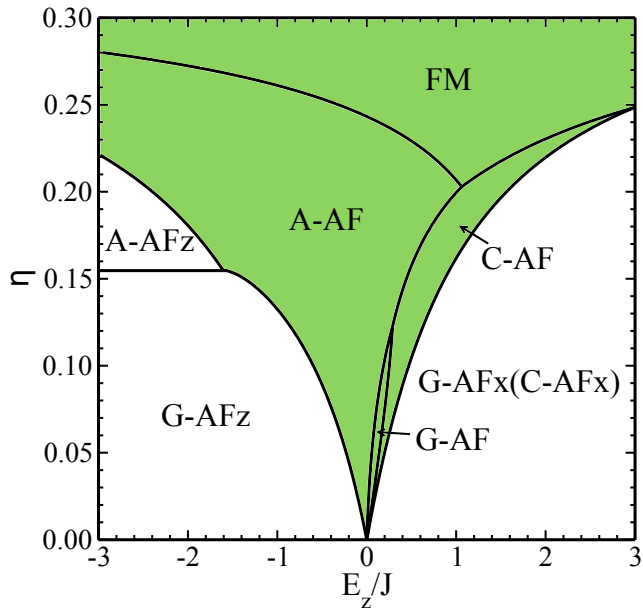}
\caption{(Color online)
Phase diagram of the 3D KK model obtained in the single-site MF
approximation. Shaded dark gray (green) area indicates phases with AO
order while the remaining magnetic phases are accompanied by FO order
with fully polarized orbitals, either $x$ (for $E_z>0$) or $z$
(for $E_z<0$). In this approach the $G$-AF$x$ and $C$-AF$x$ phases with
FO order are degenerate.}
\label{fig:3ddiag1st}
\end{figure}

The phase diagram presented in Fig. \ref{fig:3ddiag1st} was obtained
by purely energetic considerations --- it shows the border lines between
phases with the lowest energies in the $(E_{z},\eta)$ plane.
Remarkably, this phase diagram is almost the same as the one obtained
for the bilayer system\cite{wb11} --- the same phases were found both in
the 3D and in the bilayer KK model, and the major difference is the
location of the multicritical point, being now at $(E_{z},\eta)=(0,0)$.
This reflects the cubic symmetry in the model (\ref{eq:KKham}) at
$E_{z}=0$, while this symmetry is broken by the crystal-field term in
both planar models --- indeed, for the KK bilayer the multicritical
point is located at $(E_{z},\eta)=(-0.25J,0)$,\cite{wb11} and it is
moved further to $(E_{z},\eta)=(-0.5J,0)$ for the monolayer KK model.
\cite{wb12} At $\eta=0$ one finds only two AF phases:
(i) $G$-AF$z$ for $E_{z}<0$ and
(ii) $G$-AF$x$ for $E_{z}>0$, both with polarized orbital configuration
(FO order) which involves either cigar-shaped $z$ orbitals in the
$G$-AF$z$ phase, see Fig. \ref{fig:4cubes}(c), or clover-shaped $x$
orbitals in the $G$-AF$x$, see Fig. \ref{fig:4cubes}(d). Because of the
planar orbital configuration in the $G$-AF$x$ phase one finds no
interplane exchange coupling --- thus the spin order along the
$c$ axis is undetermined and this phase is degenerate with the $C$-AF
one. This degeneracy is lifted in the cluster MF approach, see below.

For higher $\eta$ the number of phases increases abruptly by those with
AO configurations (green areas), as shown in Figs. \ref{fig:4cubes}(a)
and \ref{fig:4cubes}(b), coexisting with different possible spin orders:
the $A$-AF, $C$-AF, $G$-AF and FM phases, respectively. Altogether, the
phase diagram obtained here reproduces qualitatively the one obtained
before for the simplified model using the lowest order expansion in
$\eta$.\cite{Fei97} The $C$-AF phase occurs in a narrow range of
parameters in between the $G$-AF and either $A$-AF or FM phase.

In contrast to the FO phases, the AO order in the shaded phases is never
trivial in the sense that the orbitals are never fully polarized in any
direction, which is a feature of the self-consistent MF solution
(see Ref. \onlinecite{wb11}). The only new phase with FO configuration
is the $A$-AF$z$ one appearing above $\eta=0.155$ for $E_{z}<0$. On the
other hand, the FM spin order coexists solely with alternating orbitals.
Finally, the new exotic magnetic phases reported in Sec. \ref{sec:3dcmf}
(ortho-$G$-A, canted-$A$-AF, and striped-AF phase) cannot appear here
as their stabilizing mechanism is absent in the single-site MF, so they
are not included in Table \ref{tab:sord}.

\section{Cluster mean field approach}
\label{sec:3dcmf}

\subsection{Clusters and order parameters}
\label{sec:opa}

\begin{figure}[t!]
\begin{minipage}[c]{0.5\columnwidth}%
\begin{center}
\includegraphics[clip,width=4.2cm]{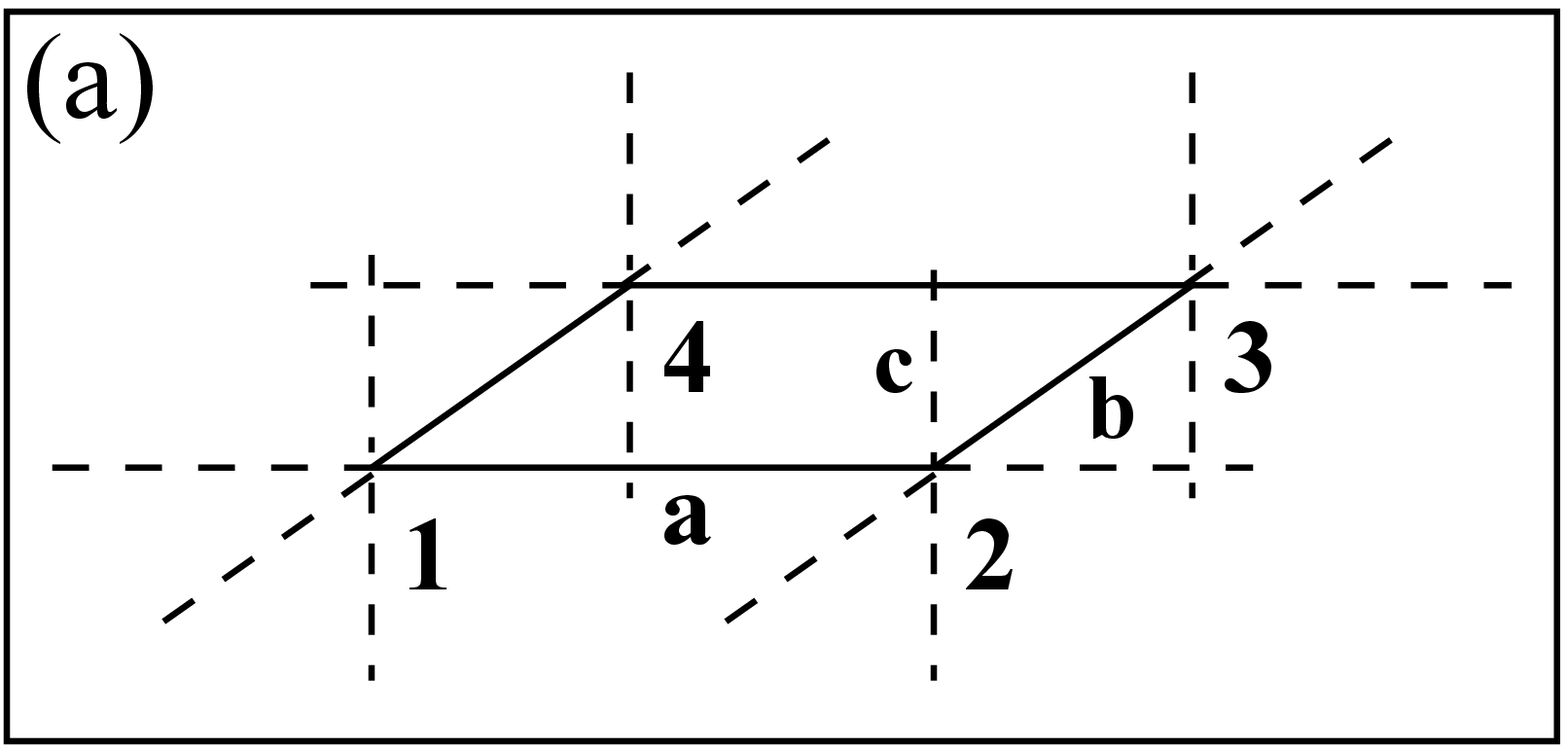}
\par\end{center}%
\end{minipage}%
\begin{minipage}[c]{0.5\columnwidth}%
\begin{center}
\includegraphics[clip,height=4.2cm]{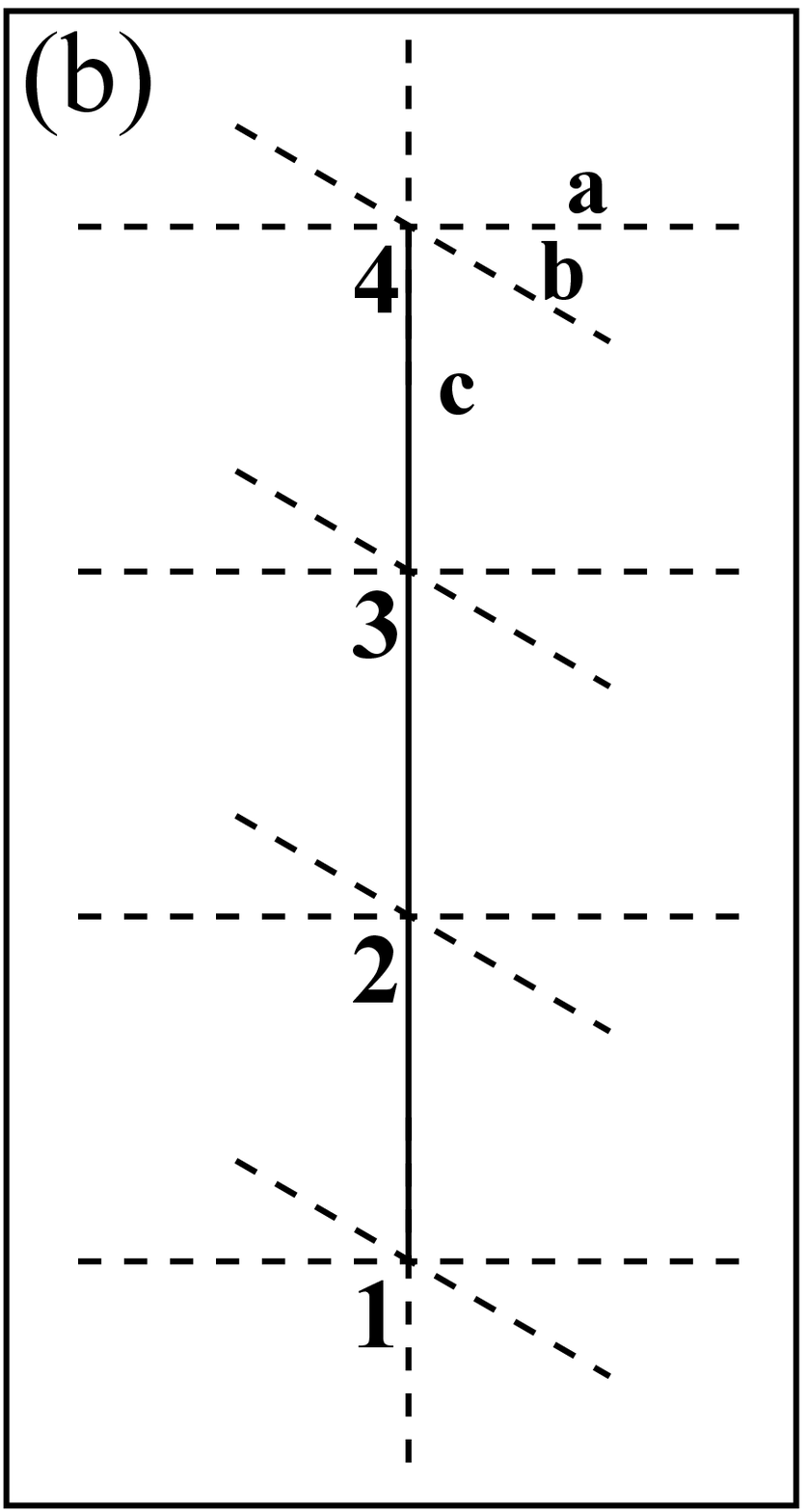}
\par\end{center}%
\end{minipage}
\caption{Schematic view of the clusters (solid lines) used in the
cluster MF approach of Sec. \ref{sec:3dcmf} to the 3D KK model:
(a) a plaquette in the $ab$ plane, and
(b) a chain along the $c$ axis.
Vertices $i=1,2,3,4$ and directions $\gamma=a,b,c$ are marked; dashed
lines stand for the outgoing bonds where the spin-orbital interactions
are replaced by the MF terms at neighboring sites,
see Eq. (\ref{eq:deco}).}
\label{fig:3dclusts}
\end{figure}

Following the ideas from Ref. \onlinecite{wb11} and \onlinecite{wb12},
to obtain more insight into the phase diagram of the 3D KK model and to
include the QF and spin-orbital entanglement on the
bonds, we divide the lattice into clusters, either plaquettes shown in
Fig. \ref{fig:3dclusts}(a), or chain clusters shown in Fig.
\ref{fig:3dclusts}(b). The most natural choice of the cluster would be a
cube with eight sites as it was done in the bilayer case\cite{wb11} but,
since we want to keep two components of the spin order parameter, we
adopted here a simpler and less time-consuming approach using four-site
clusters. The geometry of the selected clusters depends on the direction
in which we expect a large energy gain due to QF.
For example, if we want to study the transition between the $G$-AF and
$A$-AF phases then the reasonable cluster topology is a square in the
$ab$ plane, shown in Fig. \ref{fig:3dclusts}(a), because the order
along the $c$ axis does not change across the transition. On the other
hand, if we are interested in a transition between the $A$-AF and FM
phase where the order in the $ab$ planes remains constant, then a better
choice is a chain along the $c$ axis --- see Fig. \ref{fig:3dclusts}(b).

The interactions along bonds corresponding to the solid lines in Fig.
\ref{fig:3ddiag} are treated by exact diagonalization as they stand in
the 3D KK model (\ref{eq:KKham}), while the bonds represented by dashed
lines are decoupled in the MF approximation using the approximate
identity for any bond operator:
\begin{eqnarray}
O_{i}O_{j} & \approx & O_{i}\langle O_{j}\rangle
-\frac{1}{2}\langle O_{i}\rangle\langle O_{j}\rangle\nonumber \\
& + & O_{j}\langle O_{i}\rangle-\frac{1}{2}
\langle O_{i}\rangle\langle O_{j}\rangle\,.\label{eq:deco}
\end{eqnarray}
To simulate infinite 3D lattice we need MF bonds in all three
directions. Now we assume that site $i$ belongs to a chosen cluster and
$j$ belongs to a neighboring one, and the bond is splitted into two
halves --- the first one is added to the Hamiltonian of the cluster $i$
and the other one to the cluster $j$. In this way the original KK
Hamiltonian transforms into the sum of commuting cluster Hamiltonians
interacting via MF terms.

The MFs follow from Eq. (\ref{eq:deco}) applied to all the two-site
operator products encountered in the Hamiltonian (\ref{eq:KKham}) and
are defined as follows:
\begin{equation}
s_{i}^{\alpha}\equiv\left\langle S_{i}^{\alpha}\right\rangle ,
\hskip .5cm
t_{i}^{\gamma}\equiv\left\langle \tau_{i}^{\gamma}\right\rangle ,
\hskip .5cm
v_{i}^{\alpha,\gamma}\equiv
\left\langle S_{i}^{\alpha}\tau_{i}^{\gamma}\right\rangle\,.
\label{eq:mdef2}
\end{equation}
Here $\alpha=x,z$, $\gamma=a,b$ ($\gamma=c$) and $i=1,2,3,4$ for the
cluster sites of a plaquette (chain) cluster, see Fig.
\ref{fig:3dclusts}. Note that the SU$(2)$ symmetry of the spin sector
does not need to be broken in the $z$ spin direction ($\alpha=z$), but
we also allow $\alpha=x$ to capture more exotic types of magnetic order
suggested by the results reported recently for the 2D system.\cite{wb12}
However, we do not need to consider $\alpha=y$ because the KK
Hamiltonian is real.

To obtain the unbiased and the most general phase diagram, we do not
assume anything about the order {\it inside} the cluster because the
considered plaquette is small enough to keep the order parameters at
all its sites as independent variables along the MF iteration process.
Nevertheless, we still need to relate {\it different} clusters to one
another to make the problem solvable. In the case of a square cluster
we assume that the neighboring clusters in the $c$ direction can have
either the same spin or inverted spin configuration (it gives either FM
or AF bonds along $c$ axis). Furthermore, we assume
that the neighbors in the $ab$ plane can have either the same orbital
configuration that gives AO and FO orders in the $ab$ planes, or the
orbital configuration is rotated by $\pi/2$ in the $ab$ plane --- it
gives the plaquette valence-bond (PVB) phase. Similarly, we assume that
the chain clusters are copied without any change along the $c$ axis and
the neighboring chains in the $ab$ planes have:
(i) orbital configuration rotated by $\pi/2$, and
(ii) spin configuration either inverted (it gives planar AF order) or
unchanged (it gives planar FM order).
All these assumptions are necessary to solve the self-consistent cluster
MF problem and are motivated by the phase diagrams of the bilayer and
the monolayer systems (see Refs. \onlinecite{wb11} and \onlinecite{wb12}).

The self--consistency equations still cannot be solved exactly because
the effective cluster Hilbert space is of the size $d=2^{8}$ which is
too large for analytical methods. The way out is to use
Bethe--Peierls--Weiss method, i.e., to set certain initial values for
the order parameters Eq. (\ref{eq:mdef2}),
$\{s_{i}^{\alpha},t_{i}^{\gamma},v_{i}^{\alpha,\gamma}\}$, and next to
employ numerical diagonalization algorithm to the cluster Hamiltonian.
We recalculate the order parameters,
$\{s_{i}^{\alpha},t_{i}^{\gamma},v_{i}^{\alpha,\gamma}\}$,
along the iteration process, and this procedure is repeated until the
convergence conditions for energy and order parameters are satisfied.

\subsection{Phase diagram}
\label{sec:phd}

\begin{figure}[t!]
\begin{centering}
\includegraphics[clip,width=8.4cm]{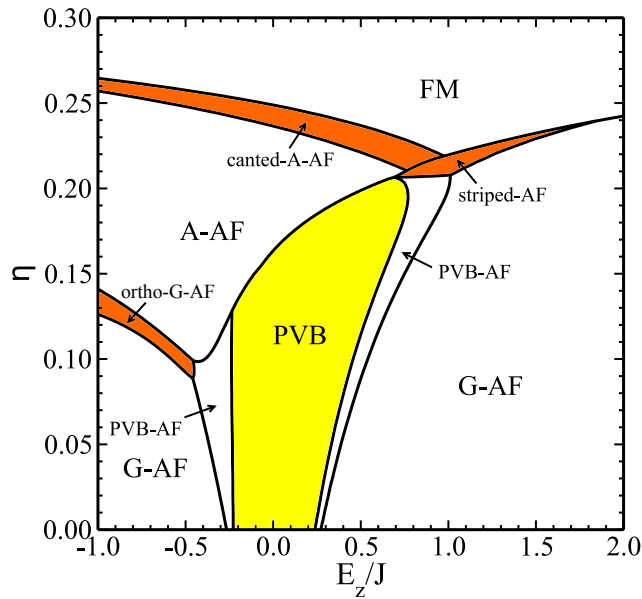}
\par\end{centering}
\caption{(Color online)
Phase diagram of the 3D KK model in the cluster MF approximation.
Plaquette valence-bond (PVB) phase with alternating spin singlets in
the $ab$ planes, highlighted in light gray (yellow), occurs between
the phases with magnetic long-range order, see Fig. \ref{fig:4cubes}.
Phases with exotic magnetic order are shaded in dark gray (orange).}
\label{fig:3ddiag}
\end{figure}

The phase diagram obtained in the cluster MF approach is shown in Fig.
\ref{fig:3ddiag}. One finds the phases with magnetic long-range order,
obtained in the single-site MF and explained in Sec. \ref{sub:1stmf_3d},
in the broad (unshaded) part of the phase diagram: the $G$-AF, $A$-AF
and FM phase. The shading in the center marks the spin disordered PVB
phase with pairs of spin singlets alternating in the $ab$ planes and
accompanied by $z$-like $3x^2-r^2/3y^2-r^2$ orbitals pointing along the
singlet bonds.\cite{Fei97} Analogous valence-bond phases were also found
in the bilayer\cite{wb11} and monolayer \cite{wb12} KK model. The darker
(orange) shading indicates exotic magnetic orders which can be found
when some AF spin interactions change into FM ones; they are:
(i) ortho-$G$-AF phase already encountered in the 2D KK model and called
there ortho-AF phase,\cite{wb12}
(ii) canted-$A$-AF phase, and
(iii) striped-AF phase.
These new phases arise from orbital fluctuations in the regimes of
strongly frustrated spin-orbital superexchange, as explained below.

\begin{figure}[t!]
\includegraphics[clip,width=5cm]{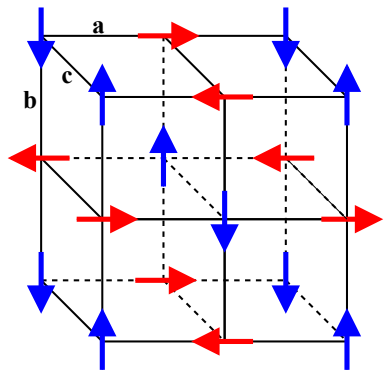}
\caption{(Color online)
Classical view of the ortho-$G$-AF spin order realized in the 3D KK
model within $ab$ planes, and staggered along the $c$ axis. Vertical and
horizontal arrows correspond to two interpenetrating AF states on the
sublattices of next nearest neighbor sites. Up (down) arrows stand for
$\left\langle S_{i}^{z}\right\rangle =\pm1/2$, right (left) arrows stand
for $\left\langle S_{i}^{x}\right\rangle =\pm1/2$. }
\label{fig:chir}
\end{figure}

We begin with the ortho-$G$-AF phase, with the spin order consisting of
two interpenetrating AF sublattices in the $ab$ planes, as shown in Fig.
\ref{fig:chir}. This configuration repeats itself in the next $ab$ plane
but all spins are inverted, meaning AF order along the $c$ axis. Note
that this phase separates phases with antiferromagnetism ($G$-AF) and
ferromagnetism ($A$-AF) within $ab$ planes, as found before in the 2D KK
model.\cite{wb12} The interactions along the $c$ axis are compatible
with the in-plane magnetic order and even stabilize it as one finds here
the ortho-$G$-AF phase at a given $\eta$ for a lower value of $E_{z}$
than that in the 2D phase. The ortho-$G$-AF phase with interplanar
antiferromagnetism replaces here the resonating valence-bond (RVB) phase
found before in the 3D KK model,\cite{Fei97} where it was proposed as
an intermediate phase separating the $G$-AF and $A$-AF phases. We
believe that the present result is more realistic (at zero temperature)
than the RVB phase within 1D chains along the $c$ axis found before
\cite{Fei97} --- this latter phase would be easily modified by any
in-plane magnetic order because the 1D Heisenberg antiferromagnet is
critical and thus easily destabilized. In addition, orbital fluctuations
remove locally AF spin coupling and thus block the resonance in the RVB
phase. We argue below that the ortho-$G$-AF phase is well justified by
the effective perturbative spin model derived for the 2D KK model in Ref.
\onlinecite{wb12} and for the present 3D model in Sec. \ref{sub:ortho}.

\begin{figure}[t!]
\includegraphics[clip,width=8.2cm]{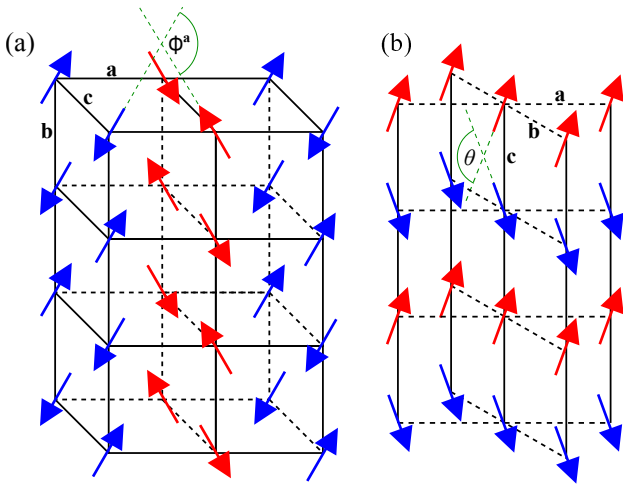}
\caption{(Color online)
Schematic views of the two exotic spin orders realized by the 3D KK
model at large Hund's exchange $\eta>0.2$:
(a) striped-AF order in the $ab$ plane, with AF order along the $c$
axis and angle $\phi^a$ between the NN spins along the $a$ axis;
(b) spin order realized in the canted-$A$-AF phase with FM order in
$ab$ planes and spin canting angle $\theta$ along the $c$ axis.
The regions of stability of these phases are shown in Fig.
\ref{fig:3ddiag} by dark gray (orange) shading.}
\label{fig:str_heli}
\end{figure}

As $\eta$ is further increased in the $A$-AF phase, one finds a second
magnetic transition, with spin correlations along the $c$ axis changing
sign. Here the canted-$A$-AF phase is found as an intermediate phase
connecting smoothly (in contrast to the ortho-$G$-AF) the $A$-AF phase
with the FM one. In the canted-$A$-AF configuration the spins are FM in
the $ab$ planes and the order along the direction $c$ changes gradually
from AF to FM with interplane spin angle $\theta$, being the canting
angle and taking values between $\theta=0$ and $\theta=\pi$, see Fig.
\ref{fig:str_heli}(b). On the other side of the phase diagram, i.e., for
$E_z>0$, one finds the striped-AF phase characterized by symmetry
breaking between the $a$ and $b$ directions in the orbital and spin
sectors, for similar values of $\eta$. The magnetic order in striped-AF
phase is AF with anisotropy; along one direction in the $ab$ plane the
order is purely AF and in the perpendicular direction the angle between
neighboring spins is close to (but not exactly) $\pi$ as shown in Fig.
\ref{fig:str_heli}(a). The orbital configuration is FO with one
preferred direction, i.e.,
$t^{a}\not=t^{b}$. Striped-AF phase connects with left $G$-AF phase by
a smooth phase transition. Further on we will present some analytical
arguments explaining both canted-$A$-AF and striped-AF phase by
perturbative expansion, see Sec. \ref{sub:cant3d} and Appendix B and
Sec. \ref{sub:stripeHs} and Appendix C, respectively.

Otherwise, the phase diagram of Fig. \ref{fig:3ddiag} contains the
$G$-AF, $A$-AF and FM configurations placed similarly as in the
single-site MF phase diagram of Fig. \ref{fig:3ddiag1st}. The degeneracy
between the left $G$-AF and $C$-AF phase is now removed, as in the
bilayer KK model,\cite{wb11} and this time we provide a perturbative
explanation of this fact in Sec. \ref{sub:3dcgaf}.
Similarly to the bilayer phase diagram, the PVB phase connects with the
left $G$-AF phase by the intermediate PVB-AF configuration, but due to
the presence of the right $G$-AF phase we have also the right PVB-AF
phase. One can summarize that the whole bottom part of the phase diagram
up to $\eta\approx0.085$ contains only smooth (second order) phase
transitions when $E_z$ is varied.
Before deriving the effective spin models for the new magnetic
configurations found in the 3D KK model we will look more closely at the
phase transitions along two cuts in the phase diagram of Fig.
\ref{fig:3ddiag}:
(i) connecting the $A$-AF and FM phases through the canted-$A$-AF phase
(Sec. \ref{sub:From--AF-to}), and
(ii) from the striped-AF to the $G$-AF phase
(Sec. \ref{sub:From-striped-AF-to}).

\subsection{From the $A$-AF to FM phase}
\label{sub:From--AF-to}

\begin{figure}[t!]
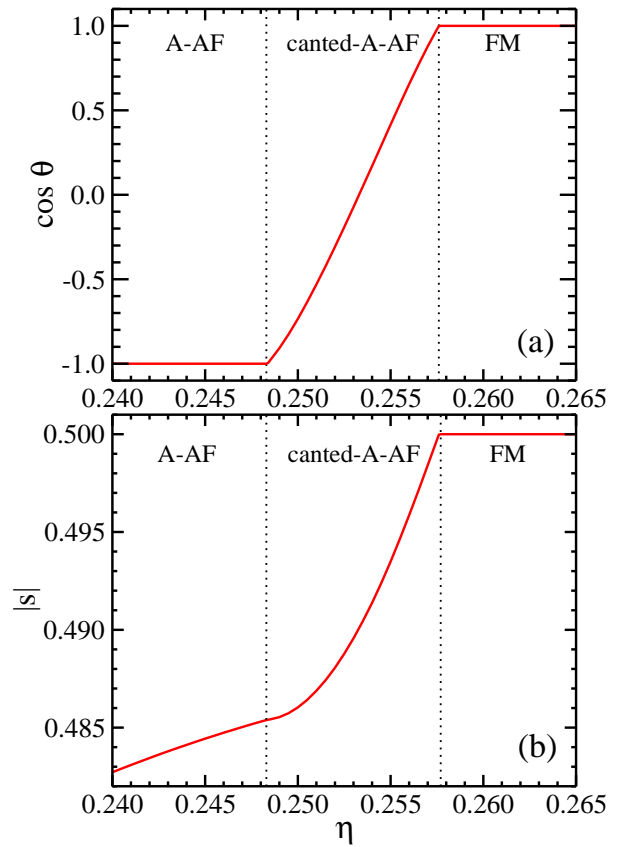

\includegraphics[clip,scale=0.5]{c2_phi}
\includegraphics[clip,scale=0.5]{c2_s}
\caption{(Color online)
Evolution of the magnetic order for increasing $\eta$ at $E_{z}=-0.5J$:
(a) cosine of the canting angle $\theta$ (\ref{eq:cos}), and
(b) total magnetization $|s|$ (\ref{eq:|s|});
both in the $A$-AF, canted-$A$-AF and FM phases from left to right.
Quantum phase transitions are indicated by vertical dotted lines.
}
\label{fig:rang_cant}
\end{figure}

First, we consider the negative crystal-field splitting $E_{z}=-0.5J$ ---
for this representative value the order changes first from the $A$-AF
into the canted-$A$-AF phase, and next into the FM phase when Hund's
exchange $\eta$ increases. We selected the chain cluster of Fig.
\ref{fig:3dclusts}(b) to study these phase transitions as the spin
order in the $ab$ planes does not change.
The changes of spin order along the $c$ axis are captured by the
cosine of the spin canting angle $\theta$ along the $c$ axis and the
total magnetization $|s|$, defined in the following way:
\begin{eqnarray}
\cos\theta&=&
\frac{1}{s^{2}}\left(s_{1}^{x}s_{2}^{x}+s_{1}^{z}s_{2}^{z}\right),
\label{eq:cos}\\
|s|&\equiv&\sqrt{\left(s^{x}\right)^{2}+\left(s^{z}\right)^{2}},
\label{eq:|s|}
\end{eqnarray}
and displayed in Figs. \ref{fig:rang_cant}(a) and \ref{fig:rang_cant}(b).
For the AF configuration (in the $A$-AF phase) one finds $\theta=\pi$
($\cos\theta=-1$), while for the FM order $\theta=0$ ($\cos\theta=1$).
In the canted-$A$-AF phase $\cos\theta$ interpolates smoothly between
these two limiting values.
Figure \ref{fig:rang_cant}(b) shows that the spin order parameter $|s|$
is gradually reduced and the QF increase when $\eta$
decreases and the $A$-AF phase is approached, but even in the $A$-AF
phase the spin order is almost classical with $|s|\simeq 0.5$. Indeed,
the QF in the $A$-AF phase with all the bonds in $ab$
planes being FM are expected to be considerably
reduced from the 2D Heisenberg antiferromagnet, as shown in the
spin-wave theory.\cite{Rac02} In the canted-$A$-AF phase the slope of
$|s|$ is the largest and the QF almost saturate when
the spins have rotated completely to the $A$-AF phase.

\begin{figure}[t!]
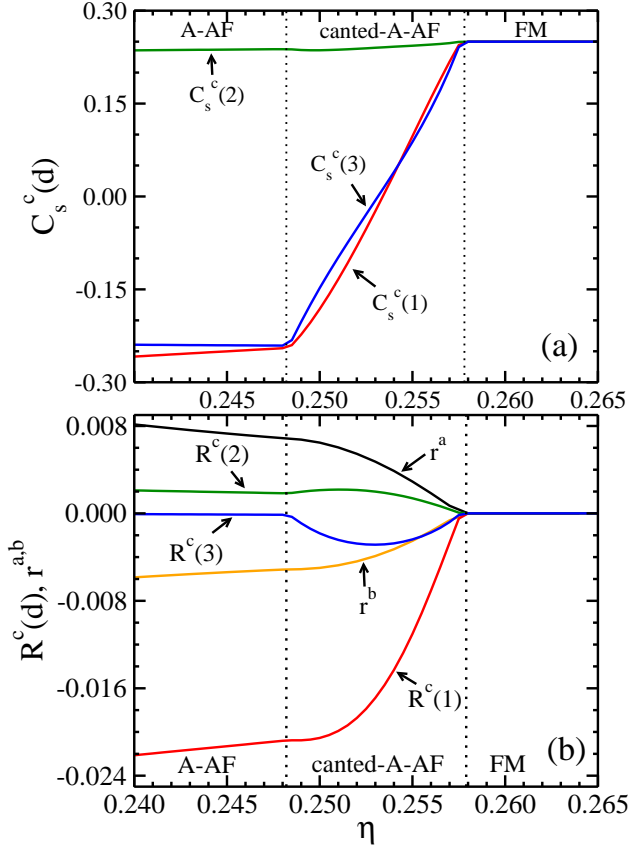

\includegraphics[clip,scale=0.5]{c2_cors}
\includegraphics[clip,scale=0.5]{c2_ent}
\caption{(Color online)
Evolution of spin and orbital correlations along the $c$ axis at
distance $d$, for $E_{z}=-0.5J$ and increasing $\eta$:
(a) spin correlation functions $C_{s}^{c}(d)$ (\ref{eq:Cs}), and
(b) on-site $r^{a(b)}$ (\ref{rsi}) and bond $R^{c}(d)$ (\ref{Rbo})
spin-orbital covariances, as obtained in the $A$-AF, canted-$A$-AF
and FM phase.
Quantum phase transitions are indicated by vertical dotted lines.
}
\label{fig:cor_ent_cant}
\end{figure}

The spin correlations,
\begin{equation}
C_{s}^{c}(d)=\langle{\bf S}_1\!\cdot\!{\bf S}_{1+d}\rangle\,,
\label{eq:Cs}
\end{equation}
along the $c$ axis for the NN, NNN and 3NN (distance $d=1,2,3$) are
shown in Fig. \ref{fig:cor_ent_cant}(a), for the same path in the
parameter space as in Fig. \ref{fig:rang_cant}. The NN and 3NN
correlations confirm that the order along the $c$ axis changes from the
AF to FM one in a continuous way, with spin correlations passing through
zero. The NNN spin correlation stays FM and is almost constant in the
entire range of $\eta$. This peculiar behavior will be explained by an
effective perturbative spin Hamiltonian in Sec. \ref{sub:cant3d}.

Figure \ref{fig:cor_ent_cant}(b) presents the spin-orbital covariances:
the on-site ones,
\begin{equation}
r_i^{\gamma}=v_i^{z,\gamma}-s_i^{z}t_i^{\gamma}\,,
\label{rsi}
\end{equation}
considered here only for the $z$ spin component $s_i^{z}$,
and the bond covariances,
\begin{equation}
R^{c}\left(d\right)=\left\langle ({\bf S}_{1}\!\cdot\!{\bf S}_{1+d})\,
\tau_{1}^{c}\tau_{1+d}^{c}\right\rangle
-\left\langle {\bf S}_{1}\!\cdot\!{\bf S}_{1+d}\right\rangle
\left\langle\tau_{1}^{c}\tau_{1+d}^{c}\right\rangle\,,
\label{Rbo}
\end{equation}
for the NN ($d=1$) and for further neighbor ($d=2,3$) operators in the
chain cluster of Fig. \ref{fig:3dclusts}(b). As one expects, in the FM
phase all the covariances vanish and the factorization of spin and
orbital operators is exact.

In the canted-$A$-AF phase the slopes of the covariances are the
steepest. The $R^{c}(1)$ function is the one of the largest magnitude
meaning that the spin-orbital entanglement on the NN bonds\cite{Ole06}
is high. In contrast to that, the covariances $R^{c}(d)$ are close to
zero for $d>1$, but $R^{c}(3)$ as the only one becomes more significant
in the canted-$A$-AF phase, indicating that this phase can be governed
by longer range spin interactions accompanied by orbital
fluctuations. The on-site covariances behave in a monotonous way and
reach relatively small absolute values meaning that they are not of
the prime importance in the considered phases.

\subsection{From the striped-AF to $G$-AF phase}
\label{sub:From-striped-AF-to}

\begin{figure}[t!]
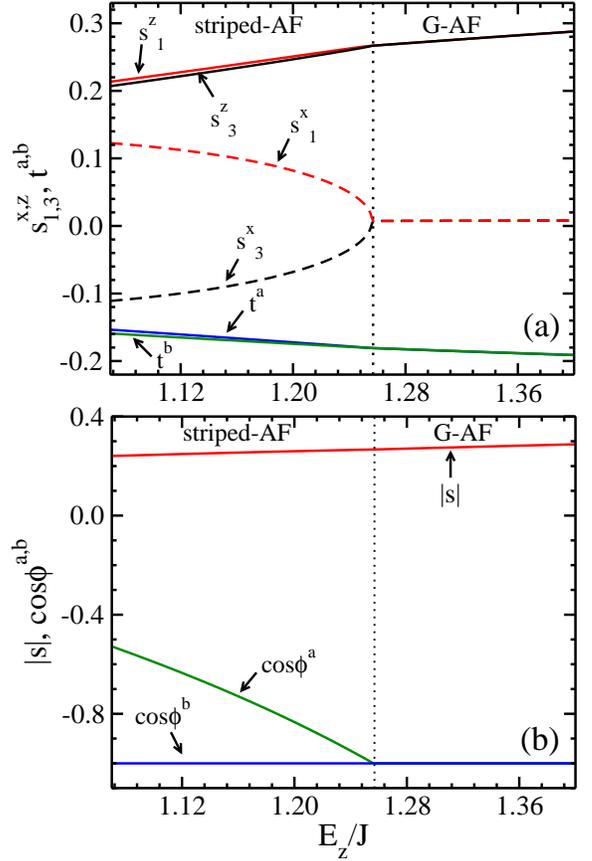

\includegraphics[clip,scale=0.5]{c3_pr}
\includegraphics[clip,scale=0.5]{c3_s_ang}
\caption{(Color online)
Evolution from the striped-AF to $G$-AF phase at $\eta=0.22$:
(a) spin $s_{1,3}^{x(z)}$ and orbital $t^{a(b)}$ order parameters, and
(b) total magnetization $|s|$ (\ref{eq:|s|}) and cosines of spin angles
$\phi^{a}$ and $\phi^{b}$, see Fig. \ref{fig:str_heli}(a).
Quantum phase transition is indicated by vertical dotted lines.
}
\label{fig:str_p_ang}
\end{figure}

Another exotic type of magnetic order found in the 3D KK model is the
striped-AF phase. This phase can evolve smoothly towards the ordinary
$G$-AF N\'eel order when $E_{z}$ increases. Here we use the plaquette
cluster of Fig. \ref{fig:3dclusts}(a) as the spin order in the $ab$
planes changes. In Fig. \ref{fig:str_p_ang}(a) we present the evolution
of the order parameters near this transition at $\eta=0.22$. The orbital
order parameters $\{t^a,t^b\}$ confirm breaking of the $a$-$b$ symmetry
in the striped-AF phase where they take slightly different values; this
difference vanishes at the phase transition. In the magnetic sector we
can distinguish four spin sublattices, see Fig. \ref{fig:str_heli}(a),
two of which are not related by a spin inversion. To show the full
complexity of the spin order we present spin averages on sites $i=1,3$
and both spin components $\alpha=x,z$. The behavior of curves confirms
the striped character of the magnetic order in the striped-AF phase,
vanishing at the transition point.

\begin{figure}[t!]
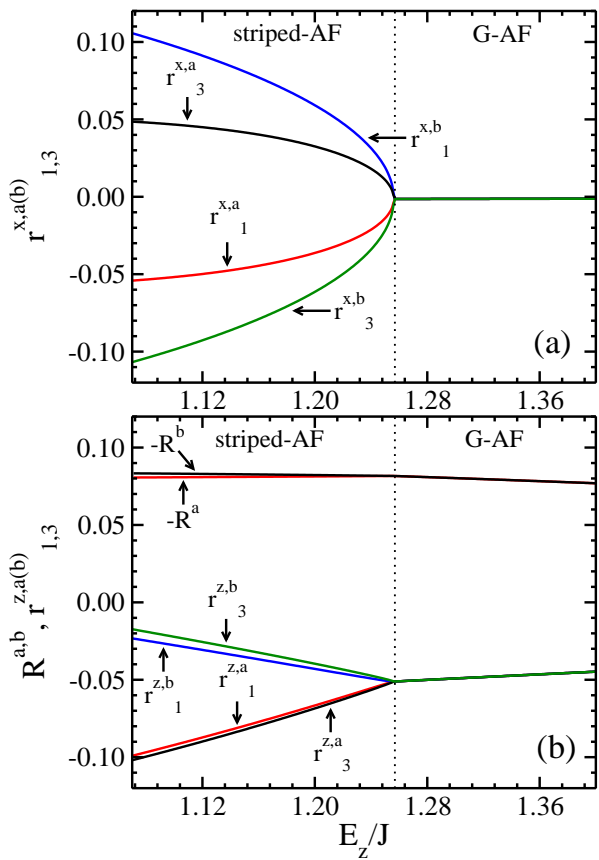

\includegraphics[clip,scale=0.5]{c3_ent1}
\includegraphics[clip,scale=0.5]{c3_ent2}
\caption{(Color online)
Evolution of covariances from the striped-AF to $G$-AF phase at
$\eta=0.22$ under increasing $E_{z}/J$ in the $ab$ planes:
(a) on-site spin-orbital covariances (\ref{rsi}) for the $x$ spin
component, $r_{1(3)}^{x,a}$ and $r_{1(3)}^{x,b}$, and
(b) on-site spin-orbital covariances (\ref{rsi}) for the $x$ spin
component, $r_{1(3)}^{z,a}$ and $r_{1(3)}^{z,b}$ (\ref{rsi}),
together with bond spin-orbital covariances $R^{a,b}$ (\ref{Rbo}).
Quantum phase transitions are indicated by vertical dotted lines.
}
\label{fig:str_ent}
\end{figure}

The quantities derived from the original order parameters, the cosines
of the angle between the neighboring spins along the $a$ and $b$ axis,
$\phi^{a(b)}$, and the total magnetization $|s|$, are shown in Fig.
\ref{fig:str_p_ang}(b). The behavior of $\cos\phi^a$ and $\cos\phi^b$
confirms the AF order along the $b$ axis independent of $E_z$, while in
the $a$ direction the angle $\phi^a$ changes at the phase transition
(at $E_z=1.27J$) from around $2\pi/3$ to $\pi$. After the transition the
cosines remain equal as expected in the isotropic AF phase. The total
magnetization $|s|$ is almost constant, increasing monotonically when
$E_z$ grows, showing that the essential physics of the striped-AF
phase lies in the spin angles, although its relatively low starting
value means that the striped-AF phase is affected by strong spin quantum
fluctuation that weaken AF order.

In Figs. \ref{fig:str_ent}(a) and \ref{fig:str_ent}(b) we present the
on-site and bond spin-orbital covariances for the same parameter range
as in Fig. \ref{fig:str_p_ang}. The bond covariance $R^{\gamma}$ for a
square cluster of Fig. \ref{fig:3dclusts}(a) is defined as
\begin{equation}
R^{\gamma}=\left\langle ({\bf S}_{1}\!\cdot\!{\bf S}_{1+\gamma})
\tau_{1}^{\gamma}\tau_{1+\gamma}^{\gamma}\right\rangle
-\left\langle {\bf S}_{1}\!\cdot\!{\bf S}_{1+\gamma}\right\rangle
\left\langle\tau_{1}^{\gamma}\tau_{1+\gamma}^{\gamma}\right\rangle,
\end{equation}
with $\gamma=a,b$. The striped-AF phase exhibits relatively large
on-site entanglement in the $x$ spin component, vanishing in the
$G$-AF phase, see Fig. \ref{fig:str_p_ang}(a). In contrast, the
entanglement in the $z$ spin component persists in the $G$-AF phase,
see Fig. \ref{fig:str_p_ang}(b). For the $x$ component the dominating
covariances are the ones for the $b$ axis and for the $z$ component
those along the $a$ axis. The fact that the $z$ covariances remain
finite in the $G$-AF phase is somewhat surprising as one could expect
that this phase with no frustration and almost fully polarized FO
configuration could be trivially factorized into spin and orbital wave
functions. This expectation based on the previous experience\cite{Ole12}
turns out to be incorrect and we show in Sec. \ref{sub:3dcgaf} that high
order orbital fluctuation are essential for stabilizing the AF order
along the $c$ axis in this phase.

\subsection{Orbital fluctuations}
\label{sec:orbif}

\begin{figure}[t!]
\begin{centering}
\includegraphics[clip,width=7.7cm]{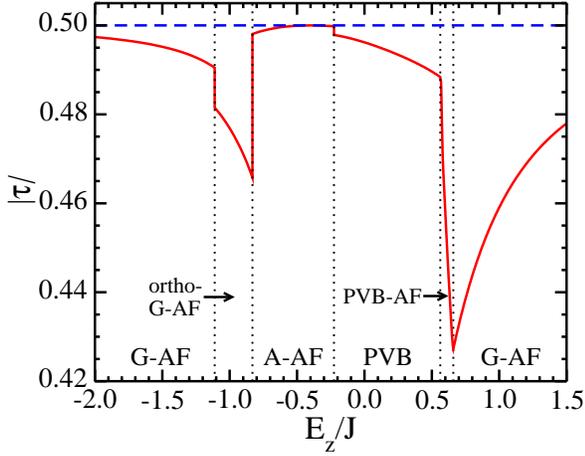}
\par\end{centering}
\caption{(Color online)
Orbital moment $|\tau|$ (\ref{eq:tau}) found in the cluster MF
approximation at the horizontal cut of the phase diagram of Fig.
\ref{fig:3ddiag} for $\eta=0.13$ and increasing $E_z$ (solid line).
Quantum fluctuations reduce $|\tau|$ from the classical value of
$0.5$ found in the single-site MF (dashed line).
Quantum phase transitions are indicated by vertical dotted lines.}
\label{fig:orbi}
\end{figure}

To estimate the strength of the orbital fluctuations in the 3D KK
model one can evaluate the total orbital moment, defined in a similar
way as the total magnetic moment of Eq. (\ref{eq:|s|}), i.e.,
\begin{equation}
|\tau|\equiv \frac{1}{2}\sqrt{\langle\sigma^z\rangle^2
+\langle\sigma^x\rangle^2}
=\frac{2\sqrt{3}}{3}\sqrt{(t^a)^2+(t^b)^2+t^at^b}.
\label{eq:tau}
\end{equation}
We investigate its value for a representative cut in the phase diagram
of Fig. \ref{fig:3ddiag}, taking $\eta=0.13$, a realistic value for
KCuF$_3$,\cite{Ole05} and for $-2J<E_z<1.5J$ within the cluster MF
and the single-site MF approximation, see Fig.
\ref{fig:orbi}. As expected, for algebraic reasons the orbital moment
(\ref{eq:tau}) in the latter approach is trivial -- $|\tau|=0.5$ for all
values of $E_z$. On the contrary, the moment $|\tau|$ found in the
cluster MF is reduced from the above maximal classical value in all
phases except for the $A$-AF phase where this
reduction is marginal. Quantum phase transitions for increasing $E_z$
are marked either by discontinuities in $|\tau|$ (first order
transitions) or by discontinuities in the derivative of $|\tau|$
(second order transitions).

The reduction of $|\tau|$ is most pronounced in the ortho-$G$-AF, where
orbital QF couple to spins, and in the PVB-AF phase where the continuous
orbital phase transition takes place but still it does not exceed $20\%$.
We observe that the orbital order in the 3D KK model is robust and
stable against weak QF in all phases. This result follows from the
rather classical directional nature of $e_g$ orbitals which leads to the
reduction of QF in the orbital space.\cite{vdB99} The orbital order
found here in the entire phase diagram justifies the perturbative
expansions in the orbital sector which are used in Sec. \ref{esm} to
derive effective spin models. We note that this case is different from
$t_{2g}$ orbitals, where orbital liquid was found both for the
perovskite lattice\cite{Kha00} and for the frustrated triangular lattice
\cite{Nor08} for the occupancy of one electron per site.

\section{Effective spin models}
\label{esm}

In this Section we describe effective spin models explaining the origin
of the exotic magnetic phases. Each model is obtained by perturbative
expansion around the ground state with orbital order stabilized by
the dominant orbital Hamiltonian ${\cal H}_0$. The expansion eliminates
the orbital degrees of freedom yielding an effective spin Hamiltonian
$H_s$ having the ground state with the exotic magnetic order.

\subsection{The ortho-$G$-AF phase}
\label{sub:ortho}

We begin with the exotic magnetic order found in the ortho-$G$-AF phase
shown in Fig. \ref{fig:chir} and derive an effective spin model for this
phase following the same ideas as those employed in the 2D KK model (see
Ref. \onlinecite{wb12}). The idea is to use the standard quantum
perturbation theory with degeneracy for the orbital sector of the KK
Hamiltonian. We can divide the Hamiltonian given by Eq. (\ref{eq:KKham})
into the unperturbed part ${\cal H}_{0}$ and perturbation ${\cal V}$ in
the following way:
\begin{eqnarray}
\label{eq:H0}
{\cal H}_{0}&\equiv&-J\varepsilon_{z}\sum_{i}\tau_{i}^{c}\,,\\
\label{eq:pert}
{\cal V}&\equiv&{\cal H}-{\cal H}_{0}\,,
\end{eqnarray}
where for simplicity we use a dimensionless parameter,
\begin{equation}
\varepsilon_z\equiv\frac{E_z}{J}.
\end{equation}
This can serve as a starting point for the perturbative treatment for
large $|E_{z}|>J$. For negative $E_{z}$ (from now on in units of $J$)
the ground state $\left|0\right\rangle$ of ${\cal H}_{0}$ is the state
with all $z$ orbitals occupied by the holes, i.e.,
\begin{equation}
\forall i:\quad\tau_{i}^{c}\left|0\right\rangle =
-\frac{1}{2}\left|0\right\rangle\,,
\end{equation}
with energy per site $\varepsilon_{0}=\frac{1}{2}J\varepsilon_z$.
Following the quantum perturbation
theory we can construct the effective spin Hamiltonian $H_{s}$ using the
expansion in powers of $\varepsilon_{z}^{-1}$:
\begin{equation}
H_{s}=N\varepsilon_{0}+\left\langle 0\right|{\cal V}\left|0\right\rangle
-\sum_{n\not=0}
\frac{|\left\langle n\right|{\cal V}\left|0\right\rangle|^2}{{\cal E}_n}
+O\left(\varepsilon_{z}^{-2}\right)\,,\label{eq:pert_exp}
\end{equation}
where all the overlaps are taken between the orbital states leaving
the spin operators alone, ${\cal E}_{n}=\left|E_{n}-E_{0}\right|$ is
the excitation energy in the physical units ($\propto J\varepsilon_z$),
and $N$ is the number of sites.
Knowing the definition of the orbital operators $\tau_{i}^{\gamma}$
Eq. (\ref{tau}), we can easily calculate the desired orbital overlaps.
The first order gives, up to the constant term,
\begin{equation}
H_{s}^{(1)}=Jg_{ab}^{(1)}\sum_{i,\gamma=a,b}
\left({\bf S}_{i}\cdot{\bf S}_{i+\gamma}\right)+Jg_{c}^{(1)}
\sum_{i}\left({\bf S}_{i}\cdot{\bf S}_{i+c}\right)\,,
\end{equation}
where $g_{ab}^{(1)}=\left(-3r_{1}+4r_{2}+r_{4}\right)/2^5$,
$g_{c}^{(1)}=\left(r_{2}+r_{4}\right)/2$ are the in-plane and
interplane coupling constants. Note that $g_{c}^{(1)}$ is positive in
the whole physical range of $\eta$, while $g_{ab}^{(1)}$ is changing
sign at $\eta_{0}\approx0.1547$. If the first order term $H_{s}^{(1)}$
alone were the only spin interaction, then $\eta_0$ would be the point
of a quantum phase transition between the $G$-AF and $A$-AF phases.
However, we have found that these two phases are separated by a stripe
of the exotic ortho-$G$-AF order.

Since $H_s^{(1)}$ vanishes at $\eta_0$ and therefore the first order
in-plane interactions can be arbitrarily weak around $\eta_{0}$, it is
necessary to go to higher order terms of the perturbative expansion to
determine the in-plane interactions leading to the exotic magnetic
order. In the second order the sum runs over all excited orbital states,
but from the superexchange terms, Eqs. (\ref{eq:KKham}) and
(\ref{eq:Hij}), one observes that ${\cal V}$ has non-zero overlap only
with states either with one or with two orbitals being excited on the
considered NN bond. This brings us to the second order correction of
the form:
\begin{equation}
H_{s,ab}^{(2)}=
\frac{Jg^{(2)}}{|\varepsilon_{z}|}
\sum_{\left\langle\left\langle ij\right\rangle\right\rangle_{ab}}\!
\left({\bf S}_{i}\cdot{\bf S}_{j}\right)-\frac{Jg^{(2)}}{2|\varepsilon_{z}|}
\sum_{\left\langle\left\langle \left\langle ij
\right\rangle\right\rangle\right\rangle_{ab}}
\!\left({\bf S}_{i}\cdot{\bf S}_{j}\right),
\label{eq:dAF_ham}
\end{equation}
with $g^{(2)}=3\left(r_{1}+2r_{2}+3r_{4}\right)^{2}/2^{10}$.
Here $\langle\langle ij\rangle\rangle_{ab}$ and
$\langle\langle\langle ij\rangle\rangle\rangle_{ab}$ denote pairs of
next NN (NNN) and third NN (3NN) sites, respectively, in the $ab$-plane
(for details of this deriviation see Appendix \ref{sec:ortho_app}). The
second order expansion also yields an $\varepsilon_{z}^{-1}$ correction
to the NN interaction strength $g_{ab}^{(1)}$, but this only moves the
transition point $\eta_{0}$ and does not change the magnetic order.

\begin{figure}[t!]
\begin{centering}
\includegraphics[clip,width=5cm]{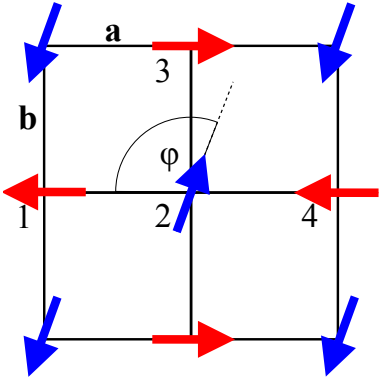}
\par\end{centering}
\caption{(Color online)
Schematic view of the order imposed by the Hamiltonian (\ref{eq:dAF_ham})
in the $ab$ plane. Interactions between NNN spins 1 and 3 (and equivalent)
are AF, between 3NN 1 and 4 are FM, but they vanish for NN spins 1 and 2,
leaving the angle $\varphi$ undetermined.}
\label{fig:chir_ang}
\end{figure}

One can easily see that the ground state of the Hamiltonian
(\ref{eq:dAF_ham}) consists of two antiferromagnets on two
interpenetrating sublattices of the square lattice, stabilized by the
NNN AF Heisenberg interaction on each sublattice. Its ground state
has long-range AF order renormalized by large QF. The
additional 3NN FM interaction makes this AF state more classical and
closer to the N\'eel order by suppressing partly QF.
In the absence of NN coupling at $\eta_0$, the two AF states on
sublattice are uncorrelated and the angle $\varphi$ between the
respective AF order parameters remains undetermined, as shown in Fig.
\ref{fig:chir_ang}.
Such ground state is similar to the ortho-$G$-AF state of Fig.
\ref{fig:chir} with AF interaction along the $c$ axis governed by
$g_{c}^{(1)}$, but we still have to find the reason why the magnetic
moments in two antiferromagnets prefer to be perpendicular to each other.

To answer this question we have to go to the third order of the
perturbative expansion given by
\begin{equation}
H_{s}^{(3)}=\sum_{n\not=0}\sum_{m\not=0}
\left\langle 0\right|{\cal V}\left|n\right\rangle\frac{1}{{\cal E}_{n}}
\left\langle n\right|{\cal V}\left|m\right\rangle\frac{1}{{\cal E}_{m}}
\left\langle m\right|{\cal V}\left|0\right\rangle .
\label{eq:h3}
\end{equation}
After long but elementary calculations, presented in more detail in
Appendix \ref{sec:ortho_app}, and under the assumption that the order
shown in Fig. \ref{fig:chir_ang} exhibits only weak QF
(see Ref. \onlinecite{wb12} for the arguments supporting this statement
in the 2D model based on the spin-wave theory, which apply even more
here in higher dimension) we obtain a classical expression for the
energy per site related with the angle $\varphi$,
\begin{equation}
\varepsilon_{\perp}^{(3)}\approx J\frac{3^{5}}{2^{21}\varepsilon_z^2}\,
(r_{1}+r_{4})\left(r_{1}\!+2r_{2}\!+3r_{4}\right)^{2}\cos^{2}\varphi\,.
\label{eq:E_perp}
\end{equation}
This expression arises from a third-order four-spin interaction. As we
can easily see, it is minimized by $\varphi=\pi/2$ which implies the
perpendicularity of the NN spins. This finally explains the origin of
the ortho-$G$-AF spin order.

\subsection{The canted-$A$-AF phase}
\label{sub:cant3d}

To explain the exotic magnetic order found in the canted-$A$-AF phase,
displayed in Fig. \ref{fig:str_heli}(b) and described in Sec.
\ref{sub:From--AF-to}, we derive an effective spin model for this phase
near the orbital degeneracy at $E_{z}=0$ in a similar way as described
above for the the ortho-$G$-AF phase. This time the starting unperturbed
orbital Hamiltonian stabilizes the AO order,
\begin{equation}
{\cal H}_{0}^x=J\varepsilon_{x}
\sum_{\gamma=a,b}\sum_{{\left\langle i,j\right\rangle \parallel\gamma}}
\sigma_{i}^{x}\sigma_{j}^{x},
\label{H0x}
\end{equation}
where we gather all the $\sigma_{i}^{x}\sigma_{j}^{x}$ terms from the
full KK Hamiltonian ${\cal H}$ in Eq. (\ref{eq:KKham})
(plus all constant terms which are not relevant and omitted here).
This sets the coupling constant in Eq. (\ref{H0x}) as
\begin{equation}
\varepsilon_{x}\equiv\frac{3}{2^7}\left(3r_{1}+r_{4}\right).
\end{equation}
The perturbation is all the rest: ${\cal V}^x={\cal H}-{\cal H}_{0}^x$.
The ground state of ${\cal H}_0^x$ is a classical configuration with
the alternating eigenstates $\sigma_{i}^{x}$. This order is consistent
with the cluster MF results showing that the canted-$A$-AF is an AO
phase, just like the neighboring $A$-AF and FM phases.

The effective spin Hamiltonian $H_{s}$ can be constructed using the
formal expansion of Eq. (\ref{eq:pert_exp}). In the zeroth order we get
the ground state energy of ${\cal H}_{0}^x$, and in the first order the
spin Hamiltonian:
\begin{equation}
H_{s}^{(1)}=-Jg_{ab}^{(1)}\sum_{i,\gamma=a,b}
\left({\bf S}_{i}\cdot{\bf S}_{i+\gamma}\right)
+Jg_{c}^{(1)}\sum_{i}\left({\bf S}_{i}\cdot{\bf S}_{i+c}\right),
\end{equation}
with $g_{ab}^{(1)}\!=7(-8r_{2}+7r_{1}-r_{4})/2^5$, $g_{c}^{(1)}\!=
(2r_{2}-r_{1}+r_{4})/2^3$. The NN interaction along the $c$ axis
$g_{c}^{(1)}(\eta)$ is changing sign at $\eta_0^x\simeq 0.236$. This
agrees well with the AF-FM crossover in the $c$ axis taking place in
the canted-$A$-AF phase that separates the $A$-AF and FM phases
for $\eta$ close to $\eta_0^x$. The NN interaction in the $ab$ planes
$g_{ab}^{(1)}$ is FM in this region. Since the first order interaction
along the $c$ axis can be made arbitrarily weak near $\eta_0^x$, we
have to go to higher orders to describe interactions between different
$ab$ planes.

In the second order ${\cal V}$ can produce two types of excited states:
(i) with a single orbital rotated by $\pi/2$ in the $ab$ plane, or
(ii) with a rotated pair of neighboring orbitals.
This leads to two type of terms in the interplanar Hamiltonian,
\begin{equation}
H_{s,c}^{(2)}=
-\frac{Jg_{c}^{(2)}}{8\varepsilon_{x}}\!\sum_{i}
\left({\bf S}_{i}\!\cdot\!{\bf S}_{i+c}\right)-
\frac{Jg_{cc}^{(2)}}{8\varepsilon_{x}}
\sum_{i}\left({\bf S}_{i-c}\!\cdot\!{\bf S}_{i+c}\right),
\label{eq:cant-2ord}
\end{equation}
with $g_{c}^{(2)}\!=\{\frac{2}{3}\left(r_{1}^{2}-r_{4}^{2}\right)
-(r_{2}+r_{4})^2\}/2^6+\frac{1}{12}\varepsilon_z(r_2+r_4)$,
and $g_{cc}^{(2)}=(r_{2}+r_{4})^{2}/2^7$ (for more details see Appendix
\ref{sec:canted_app}). The NN Heisenberg interaction $g_{c}^{(2)}$
renormalizes again the first order term $g_c^{(1)}$, modifying slightly
the value of $\eta_0^x$ where the NN coupling changes sign and makes it
dependent on $E_z$. The $E_z$-dependence of the crossover from the
$A$-AF to FM phase is consistent with the phase diagram in Fig.
\ref{fig:3ddiag}. The NNN interaction $g_{cc}^{(2)}$ is FM in agreement
with the FM NNN correlations in the canted magnetic order shown in Fig.
\ref{fig:str_heli}(b) and supported by the numerical result in Fig.
\ref{fig:cor_ent_cant}(a). The nontrivial canting angle $\theta$,
interpolating between $\theta=\pi$ and $\theta=0$ when passing from the
$A$-AF to FM phase, remains undetermined and we must proceed to the
third order of the perturbative expansion, see Appendix
\ref{sec:canted_app}.

The perturbative expansion up to third order leads to the classical
expression for the ground state energy:
\begin{equation}
\frac{E_{0}(\theta)}{J}=
\frac{\cos\theta}{4}\!\left\{ \! g_{c}^{(1)}\!
-\!\frac{1}{8\varepsilon_{x}}g_{c}^{(2)}\!+
\! O\left(\frac{1}{\varepsilon_{x}^{2}}\right)\!\right\}
\!+\!\frac{\cos^{2}\theta}{32}\frac{1}{\varepsilon_{x}^{2}}g_{c}^{(3)},
\label{eq:E_theta1}
\end{equation}
and we write
\begin{equation}
E_{0}(\theta)\equiv JA(\eta)\cos\theta+JB(\eta)\cos^{2}\theta\,.
\label{eq:E_theta}
\end{equation}
Since $B(\eta)>0$, the energy has a nontrivial minimum at $\theta_{0}$
given by the equation:
\begin{equation}
\cos\theta_{0}=
\left\{
\begin{array}{ll}
-1 & \quad {\rm for}\quad\hskip .2cm A(\eta)<-\frac12B(\eta)\\
-2\frac{A(\eta)}{B(\eta)} &
\quad {\rm for} \quad\left|A(\eta)\right|<\frac12B(\eta)\\
1 & \quad {\rm for} \quad\hskip .2cm A(\eta)>\frac12B(\eta).
\end{array}
\right.
\end{equation}
This reproduces well the behavior of $\cos\theta$ obtained via the
cluster MF method, see Fig. \ref{fig:rang_cant}(a).

\subsection{The $G$-AF versus $C$-AF spin order}
\label{sub:3dcgaf}

Here we derive an effective spin Hamiltonian around the FO ordered state
with $x$ orbitals occupied by holes to explain the energy difference
between the $G$-AF and $C$-AF phase, favoring the $G$-AF order in the
cluster MF approach, see Fig. \ref{fig:3ddiag}. As in Sec.
\ref{sub:ortho}, we can divide the Hamiltonian given by Eq.
(\ref{eq:KKham}) into the unperturbed part ${\cal H}_{0}$ (\ref{eq:H0})
and the perturbation ${\cal V}$ consists of intersite terms in the $ab$
planes (${\cal V}_{ab}$) and along the $c$ axis (${\cal V}_{c}$),
\begin{equation}
{\cal V} = J({\cal V}_{ab}+{\cal V}_{c})\,.
\label{eq:Vgaf}
\end{equation}
They are given as follows:
\begin{equation}
{\cal V}_{ab}=-\frac{1}{2}\sum_{i,\gamma=a,b}
H_{i,i+\gamma}^{\gamma}\,,\hskip .3cm
{\cal V}_{c}=-\frac{1}{2}\sum_{i}H_{i,i+c}^{c}\,.
\label{eq:Vabc}
\end{equation}
For positive $E_{z}$ the ground state $\left|0\right\rangle$ of
${\cal H}_{0}$ is the state with all $x$ orbitals occupied,
i.e.,
\begin{equation}
\forall i:\quad\tau_{i}^{c}\left|0\right\rangle
=\frac{1}{2}\left|0\right\rangle,
\label{eq:fox}
\end{equation}
with energy $E_{0}=-\frac{1}{2}J\varepsilon_{z}$ per site. Using Eq.
(\ref{eq:pert_exp}) we construct the effective spin Hamiltonian $H_{s}$
as a power expansion in ${\cal V}$. Note that the ground state
$\left|0\right\rangle$ is an eigenstate of ${\cal V}_{c}$ to zero
eigenvalue; for this reason ${\cal V}_{c}$ gives no contribution to
$H_s$ up to second order in ${\cal V}$ and the interplane interactions
appear as $\varepsilon_z^{-2}$ terms.

The effective spin Hamiltonian $H_{s}$ up to second order in ${\cal V}$
can be calculated in the same way as in Sec. \ref{sub:ortho} and
contains the same spin interactions,
\begin{eqnarray}
H_{s} & = & J\left\{g^{(1)}\!+ O\left(\varepsilon_z^{-1}\right)\!\right\}\!
\sum_{\left\langle ij\right\rangle_{ab} }\!
\left({\bf S}_{i}\!\cdot\!{\bf S}_{j}\right)+
\frac{Jg^{(2)}}{\varepsilon_{z}}\!\!\!
\sum_{\left\langle \left\langle ij\right\rangle \right\rangle_{ab} }\!\!\!
\left({\bf S}_{i}\!\cdot\!{\bf S}_{j}\right)\nonumber \\
& \!- & \frac{Jg^{(2)}}{2\varepsilon_z}\!\!
\sum_{\left\langle \left\langle \left\langle ij
\right\rangle \right\rangle \right\rangle_{ab} }
\left({\bf S}_{i}\cdot{\bf S}_{j}\right)+O\left(\varepsilon_z^{-2}\right),
\label{eq:hchir+}
\end{eqnarray}
$g^{(1)}=3(4r_{2}-r_{1}+3r_{4})/2^5$,
$g^{(2)}=3(2r_{2}+r_{4}-r_{1})^{2}/2^{10}$,
and where all bonds are in the $ab$ planes. Again, one can think of an
ortho-$G$-AF phase similar to the one obtained in the limit of
$E_{z}\to-\infty$ for $\eta$ close to $\eta_{0}$ where the NN
interaction $g^{(1)}$ vanishes (in this case $\eta_{0}\simeq0.2915$) but
this is not our aim here --- we search for an
effective description of the interplane interactions.

\begin{figure}[t!]
\begin{centering}
\includegraphics[clip,width=8.0cm]{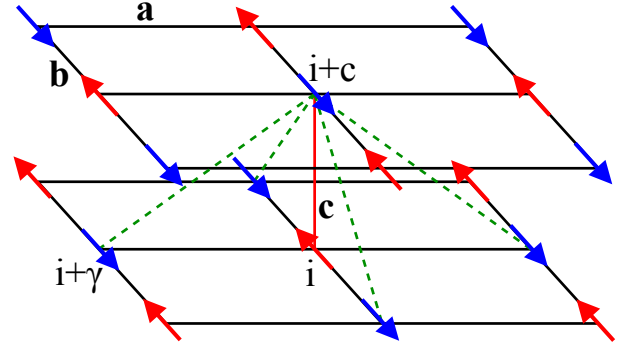}
\par\end{centering}
\caption{(Color online)
Schematic view of the $G$-AF spin order on two $ab$ planes of 3D cubic
lattice stabilized by the third-order FM NNN Heisenberg interactions
between sites $i+\gamma$ and $i+c$ shown by dashed (green) lines
(see Eq. (\ref{eq:cint})). The NN FM interactions between $i$ and $i+c$,
shown by vertical (red) line, are frustrated in the $G$-AF phase.}
\label{fig:cgaf}
\end{figure}

Looking at the third order correction to $H_{s}$ given in Eq.
(\ref{eq:h3}) and at the form of ${\cal V}_{c}$ we can see that the
interplane part of the third order
correction $H_{s,c}^{(3)}$ has the following structure:
\begin{equation}
H_{s,c}^{(3)}=\sum_{n\not=0}
\left\langle 0\right|{\cal V}_{ab}\left|n\right\rangle
\frac{1}{{\cal E}_{n}}
\left\langle n\right|{\cal V}_{c}\left|n\right\rangle
\frac{1}{{\cal E}_{n}}
\left\langle n\right|{\cal V}_{ab}\left|0\right\rangle.
\label{eq:3ord_gaf}
\end{equation}
This can be greatly simplified if we take into account two facts:
(i) ${\cal V}_{ab}$ can excite only single orbitals or pair of
neighboring orbitals so the sum over $n$ turns into the sum over
excited sites $i$, and
(ii) ${\cal V}_{c}$ is non-zero only around the excited orbitals.
The final form of $H_{s,c}^{(3)}$ can be obtained using some
identities for spin products and assuming that the spins form
classical AF state in the $ab$ planes (see Appendix \ref{sec:gaf_app}).
Of course, this is not exact because we neglect QF
but these should be small in the 3D long-range ordered
magnetic configuration being either $G$-AF or $C$-AF. The final
formula for interplane interactions is:
\begin{eqnarray}
H_{s,c}^{(3)}&=&-J\,\frac{9\delta_{r}}{2^{12}\varepsilon_{z}^{2}}
\sum_{i}\Big\{g_{c}^{(3)}\left({\bf S}_{i}\cdot{\bf S}_{i+c}\right)
\nonumber\\
&+&g_{ab}^{(3)}\sum_{\gamma}
\left({\bf S}_{i+c}\cdot{\bf S}_{i+\gamma}\right)\Big\}\,,
\label{eq:cint}
\end{eqnarray}
with $g_{c}^{(3)}=r_{1}\left(r_{1}-r_{4}\right)$ and
$g_{ab}^{(3)}=\frac{1}{4}\left(r_{1}+r_{4}\right)\left(3r_{1}-r_{4}\right)$
containing NN and NNN Heisenberg bonds both favoring ferromagnetism
($\delta_{r}$ takes only non-negative values).

This could lead to frustration as the NN bonds favor the $C$-AF state
while the NNN bonds prefer the $G$-AF one, but one can easily check that
the number of NNN bonds per site is four times larger than that of NN
bonds (see Fig. \ref{fig:cgaf}). If we take into account this factor of
$4$ comparing the coupling constants $g^{(3)}_{c}$ and $g^{(3)}_{ab}$ of
the NN and NNN interactions we obtain:
\begin{equation}
4g_{ab}^{(3)}-g_{c}^{(3)}=4\frac{1+\eta\left(1-4\eta\right)}
{\left(3\eta^{2}+2\eta-1\right)^{2}}>0
\end{equation}
in the physical range of  $\eta\in[0,1/3)$. This finally explains
why the $G$-AF phase is favored over the $C$-AF phase, as
long as we go beyond the single-site MF approximation which cannot
capture the subtle third-order orbital fluctuations (compare Figs.
\ref{fig:3ddiag1st} and \ref{fig:3ddiag}).

\subsection{The striped-AF phase}
\label{sub:stripeHs}

In Sec. \ref{sub:From-striped-AF-to} we summarized cluster MF results
for the striped-AF phase and the continuous phase transition between
the striped-AF and $G$-AF phases. These results show in particular that,
unlike all other exotic phases, the striped phase has relatively large
QF in both the orbital and the spin sector.
Nevertheless, we attempt explaining its origin by an effective spin
model keeping in mind qualitative character of our analysis.

The perturbative expansion follows the same lines as in Sec.
\ref{sub:3dcgaf}. We assume that $E_z>0$ and the unperturbed ground
state $|0\rangle$ is fully polarized by $E_z$, see Eq. (\ref{eq:fox}).
This is not an unreasonable starting approximation, because in Fig.
\ref{fig:str_p_ang}(a) we find $t^c=\langle\tau^c_i\rangle\approx 0.38$
in the striped phase, i.e., only about $12\%$ of orbitals are flipped
with respect to the fully polarized state $|0\rangle$. We believe that
it is still justified to make a perturbative expansion in the orbital
sector around the FO$x$ state. We proceed with the expansion in the same
way as in Sec. \ref{fig:str_p_ang}, but here we assume AF order along
the $c$ axis and focus on the $ab$ planes only.

Up to second order the effective Hamiltonian is given by Eq.
(\ref{eq:hchir+}) which can be rewritten in a more compact form,
\begin{eqnarray}
\label{eq:J1J2}
H_{s}
&=&
J_1\sum_{\left\langle ij\right\rangle }
\left({\bf S}_{i}\!\cdot\!{\bf S}_{j}\right)  \nonumber\\
&+&J_2
\left\{
\sum_{\left\langle \left\langle ij\right\rangle \right\rangle }
\left({\bf S}_{i}\!\cdot\!{\bf S}_{j}\right)
-\frac12
\sum_{\left\langle \left\langle \left\langle ij
\right\rangle \right\rangle \right\rangle }
\left({\bf S}_{i}\!\cdot\!{\bf S}_{j}\right)\right\},
\end{eqnarray}
with both $J_1,J_2>0$ in the physically interesting range of $\eta$ and
$E_z$. The first two terms are the same as in the $J_1-J_2$ model whose
ground state is AF when $J_1\gg J_2$ and collinear when $J_1\ll J_2$. The
extra 3NN FM coupling is consistent with the NNN AF coupling, hence the
phase diagram of the present model should be qualitatively the same.

In a classical approximation, the phase diagram of the $J_1-J_2$ model
(\ref{eq:J1J2}) motivates an Ansatz, where
\begin{eqnarray}
&&{\bf S}_i\!\cdot\!{\bf S}_{i+b}={\bf S}_i\!\cdot\!{\bf S}_{i+c}=-\frac12~,
\nonumber\\
&&{\bf S}_i\!\cdot\!{\bf S}_{i+a}=\frac12\cos\phi^a=-\frac12\cos\alpha,
\end{eqnarray}
with $\alpha=\pi-\phi^a$, see Fig. \ref{fig:str_heli}(a). Here the angle
$\alpha$ is a variational parameter. The Ansatz breaks the symmetry
between the $a$ and $b$ axes. It can describe the AF phase when
$\alpha=0$, the collinear phase when $\alpha=\pi$, and the intermediate
striped-AF phase when $\alpha\in(0,\pi)$. Up to an additive constant,
the energy per site is
\begin{equation}
\varepsilon(\alpha)~=~(-J_1+2J_2)\cos\alpha\,.
\end{equation}
It suggests a discontinuous (first order) transition between the AF and
collinear phases at $J_1=2J_2$. Thus, up to second order in the
perturbative expansion, the striped phase is unstable in the classical
version of the effective spin Hamiltonian $H_s$. However, near the
critical point, where the energy $\varepsilon(\alpha)$ does not depend
on $\alpha$, the ground state can be very susceptible
to any perturbation from higher order terms in $H_s$.

Indeed, the third order term $H^{(3)}_s$ contains the relevant
perturbation (\ref{eq:h3}) of the form
\begin{equation}
H_{s,ab}^{(3)} = J_3
\sum_{\left\langle ij\right\rangle \Vert\gamma}
\left({\bf S}_{i}\cdot{\bf S}_{j}\right)
\sum_{{\gamma'\not=\gamma\atop \gamma''\not=-\gamma}}s_{\gamma'}s_{\gamma''}
\left({\bf S}_{i+\gamma'}\cdot{\bf S}_{j+\gamma''}\right)\,,
\end{equation}
where $\gamma'$, $\gamma''$ stand for possible direction $\{a,b\}$ in
the square lattice, and $s_{\pm a(}=1$, $s_{\pm b}=-1$, respectively.
Including these interactions, the energy per site becomes
\begin{equation}
\varepsilon(\alpha)~=~(-J_1+2J_2+4J_3)\cos\alpha~+~5J_3\cos^2\alpha\,.
\end{equation}
Now the system undergoes a continuous symmetry-breaking phase transition
from the AF phase ($|\alpha|=0$) to the striped phase ($|\alpha|>0$)
when $J_1$ becomes less than $2J_2+14J_3$. This is where the
quadratic term $\varepsilon_2$ in the Landau expansion of energy, $\varepsilon(\alpha)\approx\varepsilon_0+\varepsilon_2\alpha^2
+\varepsilon_4\alpha^4$, becomes negative favoring a finite value of the
order parameter $\alpha$. The continuous character of the transition is
consistent with the cluster MF results in Fig. \ref{fig:str_p_ang}.

In conclusion, the effective classical spin model provides a qualitative
explanation for the origin of the striped-AF phase. We do not present any
quantitative predictions here --- they would be rather poor due to large
spin QF in the striped phase.

\section{Summary and conclusions}
\label{sub:summa_3d}

We have presented a rather complete analysis of the phase diagram of the
3D Kugel--Khomskii model in the framework of the cluster MF theory and
effective perturbative models. It is found that spin disordered plaquette
valence-bond phase is stable near the orbital degeneracy in the broad
range of Hund's exchange, resolving the existing controversy and in
agreement with the earlier studies.\cite{Fei97,Fei98}
Furthermore, we managed to obtain a very transparent
picture in the left part of the phase diagram of Fig. \ref{fig:3ddiag}
(for negative crystal-field splitting), with a sequence of phase
transitions occurring for increasing $\eta$ and involving two
intermediate phases with exotic magnetic orders. This sequence is caused
by the (first order) NN Heisenberg interactions changing their sign from
AF to FM spin interaction:
(i) first in the $ab$ plane, where the ortho-$G$-AF phase occurs, and
(ii) next along the $c$ axis, which leads to the canted-$A$-AF
configuration.
In both cases we found perturbative expansions around certain orbital
configurations explaining these puzzling magnetic orders by further
neighbor spin interactions. In case of the
ortho-$G$-AF phase the effective Hamiltonian is the same as for the 2D
KK model with additional AF interactions along the $c$ axis. In the
other case the intermediate configuration turned out to be essentially
classical, with FM planes damping spin quantum fluctuations, and we
have used this fact to construct the effective spin model around the AO
configuration to explain the canted-$A$-AF ordering.

To supplement these analytical considerations we presented the plots
of order parameters, correlations and spin-orbital covariances for the
two cuts in the phase diagram, passing through canted-$A$-AF and
striped-AF phases. They show that both phases involve spin-orbital
entanglement\cite{Ole12} and the plots for the canted-$A$-AF phase are
supported by the effective spin Hamiltonian derived for this phase.
The plots for the ortho-$G$-AF phase were already given in the 2D case,
\cite{wb12} and they are rather similar for the present 3D cubic lattice
(not shown).

In contrast, some of our results in the right part of the phase diagram
(for positive crystal-field splitting) are more qualitative.
As the last analytical result we showed the derivation of the effective
spin Hamiltonian for the FO$x$ configuration which explains why the
$G$-AF phase always wins over the $C$-AF order for the 3D and
bilayer systems although these two phases are degenerate in the
single-site MF approach. The answer was found in the third order of the
perturbation expansion and it was showed that the AF order along the $c$
axis is induced by the two effects:
(i) AF order in the $ab$ planes, and
(ii) NNN interaction between the planes being FM.
Surprisingly, this demonstrates that the right $G$-AF phase is highly
spin-orbital entangled, as shown in Fig. \ref{fig:str_ent} because we
need third order orbital fluctuations to stabilize the spin order,
and agrees with the plots of spin-orbital covariances presented for
the bilayer KK model.\cite{wb11} This may be the reason for a stronger
divergence of the quantum corrections to the order parameter found in
the spin-wave theory.\cite{Fei98}

Our study has shown that the striped-AF phase still requires a more
sophisticated approach than the one that was implemented here.
In particular, it is not clear why this phase appears only in the 3D
case; we have verified that it converges to a stable solution in the
2D case, but with energy being always higher than that of either the
$G$-AF or FM phase. This remains one of the open questions in the phase
diagram of the 3D Kugel--Khomskii model and further studies beyond the
cluster MF, such as the entanglement renormalization ansatz,
\cite{Vid07,*Vid08,*Cin08} are needed.

Summarizing, we have established that orbital excitations not only
couple to spin fluctuations,\cite{Woh11} but may even change the spin
order in spin-orbital systems. On the example of the 3D Kugel--Khomskii
model we have shown that three magnetic phases with exotic spin order
arise near the crossover from AF to FM spin interactions at increasing
Hund's exchange and are triggered by entangled spin-orbital quantum
fluctuations. The derived effective spin models, that include second and
third neighbor spin interactions and go beyond the Heisenberg paradigm,
provide good microscopic insights into these phases. At the same time,
these effective interactions do not confirm the earlier suggestion
\cite{Fei97} that the resonating valence-bond phase accompanied by the
FO order of $3z^2-r^2$ orbitals along the $c$ axis could be stable instead.

The derived phase diagram could in principle describe also the $A$-AF
phase observed in KCuF$_3$ at low temperature,\cite{Lak05} taking
realistic parameters, but the signatures of the 1D Heisenberg physics
are missing. On the one hand, it supports the recent view that the
Kugel-Khomskii model is incomplete and should be extended by the
Goodenough processes\cite{Ole05} and by other (lattice) degrees of
freedom\cite{Lee12} to explain fully the observed physical properties
of KCuF$_{3}$. On the other hand, the present study provides an
experimental challenge whether the ortho-$G$-AF phase could be
discovered in KCuF$_{3}$, for instance using high pressure studies.

\acknowledgments
We kindly acknowledge financial support
by the Polish National Science Center (NCN) under Projects:
No. 2012/04/A/ST3/00331 (W.B. and A.M.O.)
and
No. 2011/01/B/ST3/00512 (J.D.).

\appendix

\section{Spin model in the ortho-$G$-AF phase\label{sec:ortho_app}}

\begin{figure}[t!]
\includegraphics[clip,width=8cm]{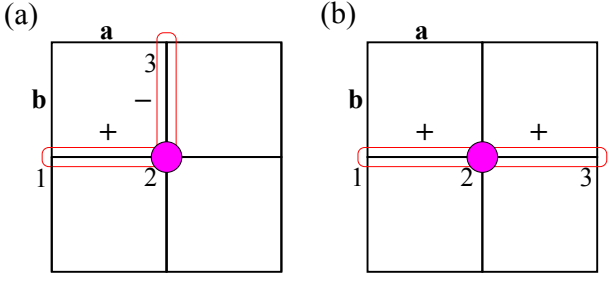}
\caption{(Color online)
Schematic views of second order corrections in the effective spin
Hamiltonian:
(a) effective NNN interactions, and
(b) effective 3NN interaction.
Red frames stand for Heisenberg bond with $\pm$ sign depending on
the bond's direction and filled (magenta) dots are the single-site
orbital excitations in the ground state.}
\label{fig:order2}
\end{figure}

The form of second order contribution to the effective in-plane
spin Hamiltonian can be derived from Eq. (\ref{eq:pert_exp}) by
calculating the matrix elements
$\left\langle n\right|{\cal V}\left|0\right\rangle$ in the orbital
sector. As a result we get
\begin{eqnarray}
\label{eq:h2}
H_{s,ab}^{(2)} & = & \frac{J}{\varepsilon_{z}}g^{(2)}
\sum_{i}\left\{ \sum_{\gamma}s_{\gamma}
\left({\bf S}_{i}\cdot{\bf S}_{i+\gamma}\right)\right\}^2 \nonumber \\
& + & \frac{J}{\varepsilon_{z}}\frac{9}{2^{11}}\!
\sum_{\left\langle ij\right\rangle }\left\{
\left({\bf S}_{i}\cdot{\bf S}_{j}\right)(r_{1}+r_{4})
+\frac{3r_{1}-r_{4}}{4}\right\}^{2}, \nonumber \\
\end{eqnarray}
where $s_{\gamma}$ is a sign factor depending on the bond's direction
$\gamma$ and originating from the definition of operators $\tau_{i}^{a(b)}$,
i.e.,
\begin{equation}
s_{\gamma}=\left\{ \begin{array}{ccc}
1 & {\rm if} & \gamma=\pm a\\
-1 & {\rm if} & \gamma=\pm b
\end{array}\right\} ,
\end{equation}
and $g^{(2)}$ is defined in Section \ref{sub:ortho}. The squared
quantities in $H_{s}^{(2)}$ produce spin products of the two forms,
shown in Fig. \ref{fig:order2}(a) and \ref{fig:order2}(b), which can
be simplified using elementary spin identities:
\begin{eqnarray}
\label{eq:Sid1}
\left({\bf S}_{i+\gamma}\cdot{\bf S}_{i}\right)
\left({\bf S}_{i}\cdot{\bf S}_{i+\gamma'}\right)&=&
\frac{1}{4}\left({\bf S}_{i+\gamma}\cdot{\bf S}_{i+\gamma'}\right)
\nonumber\\
&+&\frac{i}{2}\,{\bf S}_{i+\gamma}\!\cdot\left({\bf S}_{i}
\times{\bf S}_{i+\gamma'}\right)\,,\\
\label{eq:Sid2}
\left({\bf S}_{i}\cdot{\bf S}_{i+\gamma}\right)^{2}&=&
-\frac{1}{2}\left({\bf S}_{i}\cdot{\bf S}_{i+\gamma}\right)+\frac{3}{16}\,.
\end{eqnarray}
The latter identity simplifies the second line of $H_{s,ab}^{(2)}$ and
produces $E_{z}^{-1}$ correction to the Heisenberg interactions in
$H_{s}^{(1)}$, while the former one leads to the interactions between
further neighbors in Eq. (\ref{eq:dAF_ham}). The imaginary term in
Eq. (\ref{eq:Sid1}) is antihermitian and must cancel out with other
terms in $H_{s,ab}^{(2)}$. Thus, in $H_{s,ab}^{(2)}$, we are left with
the pure Heisenberg term
$\left({\bf S}_{i+\gamma}\cdot{\bf S}_{i+\gamma'}\right)$ connecting
sites $i+\gamma$ and $i+\gamma'$, being either NNN or 3NN, presented
in Fig. \ref{fig:order2}(a) and \ref{fig:order2}(b),
with the sign given by $s_{\gamma}s_{\gamma'}$. Due to the double
counting of the interactions the 3NN couplings in $H_{s,ab}^{(2)}$
of Eq. (\ref{eq:dAF_ham}) is twice weaker than the NNN ones.

\begin{figure}[t!]
\includegraphics[clip,width=8cm]{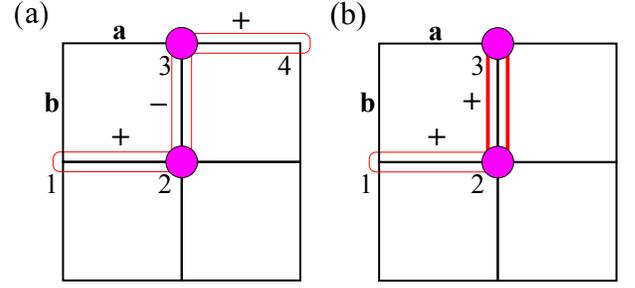}
\caption{(Color online)
Schematic view of third order corrections to $H_{s}$ (frames stand
for Heisenberg bonds with $\pm$ sign depending on their direction,
filled circles indicate orbital flips):
(a) term bringing new physics, chain product of three bonds,
(b) term doubling the results from second order expansion, interaction
between spins $2$ and $3$ is squared (bond marked by a thick frame).}
\label{fig:order3}
\end{figure}

The third order correction necessary to determine the in-plane NN
interaction at $\eta_{0}$ is given by Eq. (\ref{eq:h3});
here we limit the sums over the excited orbital states
$\left\{\left|n\right\rangle,\left|m\right\rangle\right\}$
only to the ones with orbital flips lying in the same $ab$ plane.
This produces many contributions to the spin Hamiltonian but we are
interested only in terms with new operator structure with respect to
lower orders because the others will be just the $E_{z}^{-2}$
corrections to the already existing interactions. The terms bringing
potentially new physics are the ones with three different Heisenberg
bonds multiplied one after another. Such contribution is depicted in
Fig. \ref{fig:order3}(a) for sites $i=1,2,3,4$ (in contrast, Fig.
\ref{fig:order3}(b) shows the term bringing no new contribution to
$H_{s}$) and can be transformed following the identity,
\begin{eqnarray}
& &\left({\bf S}_{1}\!\cdot\!{\bf S}_{2}\right)
\left({\bf S}_{2}\!\cdot\!{\bf S}_{3}\right)
\left({\bf S}_{3}\!\cdot\!{\bf S}_{4}\right)=
\nonumber \\
&   & \frac{1}{16}\,{\bf S}_{1}\!\cdot\!{\bf S}_{4}+\frac{1}{4}
\left({\bf S}_{1}\!\cdot\!{\bf S}_{4}\right)
\left({\bf S}_{2}\!\cdot\!{\bf S}_{3}\right)
\nonumber \\
& - &\frac{1}{4}\left({\bf S}_{1}\!\cdot\!{\bf S}_{3}\right)
\left({\bf S}_{2}\!\cdot\!{\bf S}_{4}\right)
+ \frac{i}{8}\,{\bf S}_{1}\!\cdot\left({\bf S}_{3}\times{\bf S}_{4}\right)
\nonumber \\
& + &\frac{i}{8}\,{\bf S}_{1}\!\cdot\left({\bf S}_{4}
\times{\bf S}_{2}\right)
+ \frac{i}{8}\,{\bf S}_{1}\!\cdot\left({\bf S}_{4}\times{\bf S}_{2}\right),
\label{eq:3heis}
\end{eqnarray}
where the cross-product terms are antihermitian and must cancel out with
other terms of the same structure in $H_{s,ab}^{(3)}$. To analyze the
remaining terms in Eq. (\ref{eq:3heis}) it is helpful to employ the
almost classical nature of the AF spin order on two sublattices:
(i) the first term is an AF interaction between the sublattices which is
{\it not} compatible with the antiferromagnetism on sublattices that is
one order of $E_{z}$ stronger,
(ii) the second term favors perpendicularity of the two AF sublattices,
as long as its sign is positive, which is compatible with the order on
sublattices, and
(iii) the third term brings no new information about the spin order.

Putting all together, we can argue that the third order perturbative
contributions in Eq. (\ref{eq:3heis}) can favor perpendicularity of the
order parameters on two sublattice antiferromagnets. Now we have to
extract all such contributions from Eq. (\ref{eq:h3})and check
whether the total sign is indeed positive because otherwise our
arguments would be incomplete. After lengthy but elementary calculation
(see Ref. \onlinecite{wb12} for more details) we obtain the
third order Hamiltonian, with interactions not included in lower orders
of the form:
\begin{eqnarray}
H_{s,ab}^{(3)}&=&J\,\frac{27}{2^{17}\,
\varepsilon_{z}^{2}}\left(r_{1}+r_{4}\right)
\left(r_{1}+2r_{2}+3r_{4}\right)^{2} \nonumber\\
 & \times & \sum_{\left\langle ij\right\rangle \Vert\gamma}
 \left({\bf S}_{i}\cdot{\bf S}_{j}\right)\sum_{{\gamma'\not=
\gamma\atop \gamma''\not=-\gamma}}s_{\gamma'}s_{\gamma''}
 \left({\bf S}_{i+\gamma'}\!\cdot\!{\bf S}_{j+\gamma''}\right).\nonumber\\
\end{eqnarray}
These are the special contributions of the type shown in Fig.
\ref{fig:order3}(a) taken into account here. This, in the classical
limit, gives the energy $\varepsilon_{\perp}^{(3)}$ in Eq.
(\ref{eq:E_perp}) favoring perpendicular orientation of the NN spins.

\section{Spin model in the canted-$A$-AF phase\label{sec:canted_app}}

To calculate the second order correction to the spin Hamiltonian $H_{s}$
in the canted-$A$-AF phase we have to, as before, calculate the matrix
element (overlap) $\left\langle n\right|{\cal V}^x\left|0\right\rangle$
in the orbital sector. Since the state $\left|n\right\rangle $ is an
excited state of ${\cal H}_0$, the non-zero overlap can be obtained only
for the part ${\cal V}^x$ of the interaction ${\cal V}$ that contains
at least one operator $\sigma_{i}^{z}$. Assuming FM order
in the $ab$ planes it can be written as,
\begin{eqnarray}
{\cal V}^x & =&
\frac{Jr_{1}}{2^{5}}\sum_{{\left\langle ij\right\rangle\parallel\atop\gamma=a,b}}
\left\{\sigma_{j}^{z}\sigma_{i}^{z}-s_{\gamma}\sqrt{3}
\left(\sigma_{j}^{z}\sigma_{i}^{x}+\sigma_{j}^{x}\sigma_{i}^{z}\right)
\right\} \nonumber \\
& + & \frac{J}{8}\sum_{{\left\langle ij\right\rangle \parallel\atop c}}
\left\{ \left( r_{1}\Pi_{t}^{(ij)}-r_{4}
\Pi_{s}^{(ij)}\right)\sigma_{i}^{z}\sigma_{j}^{z}\right.\nonumber \\
&+&\left.(r_{2}+r_{4})\Pi_{s}^{(ij)}\left(\sigma_{i}^{z}
+\sigma_{j}^{z}\right)\right\}
-\frac{1}{2}\,E_{z}\sum_{i}\sigma_{i}^{z}\,.
\end{eqnarray}
employing the spin projectors in Eqs. (\ref{eq:proje}).
Now we can derive the second order Hamiltonian,
\begin{eqnarray}
H_{s}^{(2)} & = & -\frac{J}{8\varepsilon_{x}}\sum_{i}
\left\{ \frac{\varepsilon_{z}}{2}+\frac{r_{2}+ r_{4}}{8}
\left(\frac{1}{2}-\sum_{\gamma=\pm c}
{\bf S}_{i}\!\cdot\!{\bf S}_{i+\gamma}\right)\right\}^{2}
\nonumber \\
 & - & \frac{J}{3\varepsilon_{x}}\sum_{i}
 \frac{1}{2^{8}}\left\{ (r_{1}+ r_{4})({\bf S}_{i}\cdot{\bf S}_{i+c})
 +\frac{3r_{1}- r_{4}}{4}\right\}^{2},
\nonumber \\
\end{eqnarray}
which can be easily transformed into Eq. (\ref{eq:cant-2ord}) using
the identity (\ref{eq:Sid1}).

In the third order we are interested only in terms with three Heisenberg
bonds multiplied by each other along the $c$ direction,
according to Eq. (\ref{eq:3heis}), as all other possible products will
only reproduce the results from lower orders. Such terms
lead to the correction $H_{s,c}^{(3)}$ of the form:
\begin{equation}
H_{s,c}^{(3)}\!=\frac{Jg_{c}^{(3)}}{\varepsilon_{x}^{2}}\!\sum_{i}\!\left\{
\!\left({\bf S}_{i+c}\!\cdot\!{\bf S}_{i}\right)
\!\left({\bf S}_{i}\!\cdot\!{\bf S}_{i-c}\right)
\!\left({\bf S}_{i-c}\!\cdot\!{\bf S}_{i-2c}\right)+ {\rm H.c.}\right\},
\end{equation}
with $g_c^{(3)}=\frac{5}{6}\left(r_2+r_4\right)^2\left(r_4+r_1\right)/2^{14}$.
This can be further simplified using the identity (\ref{eq:Sid2}),
\begin{eqnarray}
H_{s,c}^{(3)} & \!=\! & J\frac{g_{c}^{(3)}}{8\varepsilon_{x}^{2}}\!
\sum_{i}\left\{
\left({\bf S}_{i+c}\!\cdot\!{\bf S}_{i-2c}\right)
+4\left({\bf S}_{i+c}\!\cdot\!{\bf S}_{i-2c}\right)
\!\left({\bf S}_{i}\!\cdot\!{\bf S}_{i-c}\right)\right.\nonumber \\
& \!-\! & \left. 4\left({\bf S}_{i+c}\!\cdot\!{\bf S}_{i-c}\right)
 \left({\bf S}_{i}\!\cdot\!{\bf S}_{i-2c}\right)\right\} .
 \label{eq:cant_3ord}
\end{eqnarray}
Now, according to Fig. \ref{fig:str_heli}(b), we use the classical
expressions for the spin scalar product using the canting angle $\theta$
for the odd neighbors and FM order for the even neighbors, imposed
by the second order Hamiltonian of Eq. (\ref{eq:cant-2ord}), i.e.,
\begin{equation}
{\bf S}_{i}\cdot{\bf S}_{i+(2n-1)c}=\frac{1}{4}\cos\theta,\hskip.5cm
{\bf S}_{i}\cdot{\bf S}_{i+2nc}=\frac{1}{4}.
\end{equation}
After inserting this into Eq. (\ref{eq:cant_3ord}) only the first
line depends on $\theta$ and gives a contribution to the classical
ground state energy $E_{0}(\theta)$ of the canted-$A$-AF phase see
Eq. (\ref{eq:E_theta}).

\section{Spin model in the $G$-AF phase\label{sec:gaf_app}}

Here we consider the $G$-AF phase with FO$x$ order for $E_z>0$.
The third order contribution of the form given by Eq. (\ref{eq:3ord_gaf})
can be expressed as,
\begin{eqnarray}
H_{s,c}^{(3)} & \!= & -J\frac{g^{(2)}\delta_{r}}{2\varepsilon_{z}^{2}}\!
\sum_{i,\gamma,\gamma'}\! s_{\gamma}s_{\gamma'}
\left({\bf S}_{i}\!\cdot\!{\bf S}_{i+\gamma}\right)
\left({\bf S}_{i}\!\cdot\!{\bf S}_{i+c}\right)
\left({\bf S}_{i}\!\cdot\!{\bf S}_{i+\gamma'}\right)\nonumber \\
& \!- &J\frac{9\delta_{r}}{2^{13}\varepsilon_{z}^{2}}
\sum_{i,\gamma}\left\{(r_{1}+r_{4})
\left({\bf S}_{i}\!\cdot\!{\bf S}_{i+\gamma}\right)
+\frac{3r_{1}-r_{4}}{4}\right\} \nonumber \\
& \!\times & \left({\bf S}_{i}\!\cdot\!{\bf S}_{i+c}\right)
\left\{(r_{1}+r_{4})\left({\bf S}_{i}\!\cdot\!{\bf S}_{i+\gamma}\right)
+\frac{3r_{1}-r_{4}}{4}\right\},
\label{eq:h3_gaf}
\end{eqnarray}
where $\delta_{r}=r_{1}-r_{2}$, sums over $\gamma$ and $\gamma'$ run
over the in-plane directions $\left\{ \pm a,\pm b\right\}$. The first
term in Eq. (\ref{eq:h3_gaf}) comes from single-orbital excitations,
while the second one (second and third line) --- from two-orbital
excitations.
The interplane interactions are typically sandwiched between two
in-plane bonds; these can be treated with the following spin identity:
\begin{eqnarray}
\!&&\left({\bf S}_{i}\!\cdot\!{\bf S}_{i+\gamma}\right)
\left({\bf S}_{i}\!\cdot\!{\bf S}_{i+c}\right)
\left({\bf S}_{i}\!\cdot\!{\bf S}_{i+\gamma'}\right)  \nonumber \\
\!&=&\frac{1}{4}\left({\bf S}_{i+c}\!\cdot\!{\bf S}_{i+\gamma}\!\right)
\left({\bf S}_{i}\!\cdot\!{\bf S}_{i+\gamma'}\!\right)
-\frac{1}{4}\left({\bf S}_{i+\gamma'}\!\cdot\!{\bf S}_{i+\gamma}\right)
\left({\bf S}_{i}\!\cdot\!{\bf S}_{i+c}\right) \nonumber \\
\!& + & \frac{1}{4}\left({\bf S}_{i}\!\cdot\!{\bf S}_{i+\gamma}\!\right)
\left({\bf S}_{i+c}\!\cdot\!{\bf S}_{i+\gamma'}\!\right)
+\frac{i}{8}\,{\bf S}_{i+\gamma}\!\cdot\!\left({\bf S}_{i+c}
\times{\bf S}_{i+\gamma'}\right)\,.\nonumber \\
\label{eq:Sid3}
\end{eqnarray}
Under the assumption of the classical AF order in $ab$ planes, i.e.,
$\left({\bf S}_{i}\cdot{\bf S}_{i+\gamma}\right)\equiv-\frac{1}{4}$ for
$\gamma=\pm a,\pm b$, the above identity gives the following results
for the tri-quadratic spin products in $H_{s,c}^{(3)}$,
\begin{eqnarray}
&\sum_{\gamma,\gamma''}&s_{\gamma}s_{\gamma''}
\left({\bf S}_{i}\!\cdot\!{\bf S}_{i+\gamma}\right)
\left({\bf S}_{i}\!\cdot\!{\bf S}_{i+c}\right)
\left({\bf S}_{i}\!\cdot\!{\bf S}_{i+\gamma''}\right)\nonumber \\
&=&\frac{1}{4}\sum_{\gamma}\left({\bf S}_{i+c}\!\cdot\!{\bf S}_{i+\gamma}\right)
+\frac{1}{4}\left({\bf S}_{i}\!\cdot\!{\bf S}_{i+c}\right)\,,
\end{eqnarray}
and
\begin{equation}
\left({\bf S}_{i}\!\cdot\!{\bf S}_{i+\gamma}\right)
\left({\bf S}_{i}\!\cdot\!{\bf S}_{i+c}\right)
\left({\bf S}_{i}\!\cdot\!{\bf S}_{i+\gamma}\right)=
-\frac{1}{16}\left({\bf S}_{i}\!\cdot\!{\bf S}_{i+c}\right)\,.
\end{equation}
The biquadratic terms can be simplified using identity (\ref{eq:Sid1}).
Note that the classical AF order,
$\left({\bf S}_{i}\cdot{\bf S}_{i+\gamma}\right)\equiv-\frac{1}{4}$
for $\gamma=\pm a,\pm b$, is inserted {\it after} the spin products
are sorted out with spin identities (\ref{eq:Sid1}) and (\ref{eq:Sid3})
but not before. This is an important distinction:
we extract the result of higher order in-plane spin fluctuations before
we freeze them out to get a simplified picture.
After gathering all the interactions together we obtain $H_{s,c}^{(3)}$
in Eq. (\ref{eq:cint}).

\bibliographystyle{apsrev4-1}

\end{document}